\documentclass[prb,aps,twocolumn,superscriptaddress,longbibliography]{revtex4-2}

\usepackage{graphicx,color}
\usepackage{amsthm}
\usepackage{amsfonts}
\usepackage{algorithmic}
\usepackage{enumerate}
\usepackage{latexsym}
\usepackage{amsmath}
\usepackage{amssymb}
\usepackage{ulem}
\usepackage[dvipsnames,svgnames,x11names]{xcolor}
\usepackage[colorlinks=true,citecolor=blue,linkcolor=blue]{hyperref}
\usepackage[T1]{fontenc}
\usepackage[english]{babel}
\usepackage{diagbox}

\def\avg#1{\langle#1\rangle}

\begin{document}
\title{Quantum highway: Observation of minimal and maximal speed limits  for few and many-body states}
\author{Zitian Zhu}
\thanks{These authors contributed equally}
\affiliation{School of Physics, Zhejiang University, Hangzhou 310027, China}
\affiliation{ZJU-Hangzhou Global Scientific and Technological Innovation Center, Zhejiang University, Hangzhou 311215, China}

\author{Lei Gao}
\thanks{These authors contributed equally}
\affiliation{Beijing Computational Science Research Center, Beijing 100193, China}

\author{Zehang Bao}
\thanks{These authors contributed equally}
\affiliation{School of Physics, Zhejiang University, Hangzhou 310027, China}
\affiliation{ZJU-Hangzhou Global Scientific and Technological Innovation Center, Zhejiang University, Hangzhou 311215, China}

\author{Liang Xiang}
\author{Zixuan Song}
\author{Shibo Xu}
\author{Ke Wang}
\author{Jiachen Chen}
\author{Feitong Jin}
\author{Xuhao Zhu}
\author{Yu Gao}
\author{Yaozu Wu}
\author{Chuanyu Zhang}
\author{Ning Wang}
\author{Yiren Zou}
\author{Ziqi Tan}
\author{Aosai Zhang}
\author{Zhengyi Cui}
\author{Fanhao Shen}
\author{Jiarun Zhong}
\author{Tingting Li}
\author{Jinfeng Deng}
\author{Xu Zhang}
\author{Hang Dong}
\author{Pengfei Zhang}
\affiliation{School of Physics, Zhejiang University, Hangzhou 310027, China}
\affiliation{ZJU-Hangzhou Global Scientific and Technological Innovation Center, Zhejiang University, Hangzhou 311215, China}
\author{Zhen Wang}
\author{Chao Song}
\affiliation{School of Physics, Zhejiang University, Hangzhou 310027, China}
\affiliation{ZJU-Hangzhou Global Scientific and Technological Innovation Center, Zhejiang University, Hangzhou 311215, China}
\affiliation{Hefei National Laboratory, Hefei 230088, China}

\author{Chen Cheng}
\affiliation{Key Laboratory of Quantum Theory and Applications of MoE, Lanzhou Center for Theoretical Physics, and Key Laboratory of Theoretical Physics of Gansu Province, Lanzhou University, Lanzhou, Gansu 730000, China}

\author{Qiujiang Guo}
\email{qguo@zju.edu.cn}
\affiliation{School of Physics, Zhejiang University, Hangzhou 310027, China}
\affiliation{ZJU-Hangzhou Global Scientific and Technological Innovation Center, Zhejiang University, Hangzhou 311215, China}
\affiliation{Hefei National Laboratory, Hefei 230088, China}

\author{Hekang Li}
\email{hkli@zju.edu.cn}
\affiliation{School of Physics, Zhejiang University, Hangzhou 310027, China}
\affiliation{ZJU-Hangzhou Global Scientific and Technological Innovation Center, Zhejiang University, Hangzhou 311215, China}

\author{H. Wang}
\affiliation{School of Physics, Zhejiang University, Hangzhou 310027, China}
\affiliation{ZJU-Hangzhou Global Scientific and Technological Innovation Center, Zhejiang University, Hangzhou 311215, China}
\affiliation{Hefei National Laboratory, Hefei 230088, China}

\author{Haiqing Lin}
\affiliation{School of Physics, Zhejiang University, Hangzhou 310027, China}
\affiliation{Beijing Computational Science Research Center, Beijing 100193, China}

\author{Rubem Mondaini}
\email{rmondaini@uh.edu}
\affiliation{Department of Physics, University of Houston, Houston, Texas 77004, USA}
\affiliation{Texas Center for Superconductivity, University of Houston, Houston, Texas 77204, USA}


\begin{abstract}
Tracking the time evolution of a quantum state allows one to verify the thermalization rate or the propagation speed of correlations in generic quantum systems. Inspired by the energy-time uncertainty principle, bounds have been demonstrated on the maximal speed at which a quantum state can change, resulting in immediate and practical tasks. Based on a programmable superconducting quantum processor, we test the dynamics of various emulated quantum mechanical systems encompassing single- and many-body states. We show that one can test the known quantum speed limits and that modifying a single Hamiltonian parameter allows the observation of the crossover of the different bounds on the dynamics. We also unveil the observation of \textit{minimal} quantum speed limits in addition to more common maximal ones, i.e., the lowest rate of change of a unitarily evolved quantum state. Our results establish a comprehensive experimental characterization of quantum speed limits and pave the way for their subsequent study in engineered non-unitary conditions.
\end{abstract}

\maketitle


\section*{Introduction \label{sec:introduction}}
\indent 
The central question of quantum non-equilibrium dynamics revolves around exploring how quickly a quantum state can be modified~\cite{Eisert_2015_nonequilibrium}. It presents an elemental step towards understanding the rate of equilibration and thermalization of generic quantum systems~\cite{Rigol2008}, the propagation speed of information~\cite{Lieb1972_LR_Bounds, Cheneau2012_LR_Bounds_Exp} as well as practical applications, such as quantum state transfer~\cite{Bose_2003_QST, Xiang_2024_QST}, the minimal time of applying quantum gates~\cite{Ashhab_2012_theory}, and the analysis of optimal quantum control~\cite{Krotov1996_optimal_control_book, T_2009_PRL_OptimalControlQSL}. When putting the interpretation of the energy-time uncertainty relation~\cite{Heisenberg1927} on a firmer ground~\cite{T_2017_QSLreview}, Mandelstam and Tamm (MT)~\cite{T_1945_JPUSSR_OriginalMTbound} realized that it sets the fundamental bound of the intrinsic time scale on how fast a unitary quantum dynamics can evolve.

Such bound constrains the unitary process of a (pure or mixed) initial state to its \textit{orthogonal} counterpart, defining a minimal evolution time $t_\perp$ obeying 
$t_\perp\geq t_{\rm MT}\equiv \pi\hbar/(2\Delta E)$, where $\Delta E=\left(\avg {\hat{H}^2}-\avg {\hat{H}}^2\right)^{1/2}$ is the energy spread of the time-independent Hamiltonian $\hat{H}$ governing the system and $\avg {\hat H} \equiv E$ is the mean energy. Building on that, Margolus and Levitin (ML)~\cite{T_1998_PhysicaD_MLbound} further demonstrated that $E$ with respect to the ground state energy $E_{\rm min}$ establishes a second orthogonalization bound, $t_\perp\geq t_{\rm ML}\equiv \pi\hbar/[2(E-E_{\rm min})]$. Together, they construct the `quantum speed limit' (QSL) concept, defining the lower bounds of the time required for the unitary dynamics of \textit{any} quantum system to unfold. Substantial investigation across various domains, encompassing both isolated and open systems, unitary and non-unitary evolutions, and spanning the realms of fundamental  
~\cite{T_2003_PRA_EntanglementSpeedupQSL,T_2006_PRA_GeneralizedML,T_2012_PRA_GeneralizedMTFisher,T_2013_PRL_NonunitaryFisherQSL,T_2013_PRL_QSLOpenSystem,T_2013_PRL_QSLNon-Markovian,T_2014_SciRep_FanHeng,T_2015_SciRep_QSLNon-Markovian,T_2016_PRX_GeneralizedGeometricQSL,T_2018_PRL_QSLAllStates,T_2018_PRL_QSLQuantum-to-ClassicalTransition,T_2019_PRL_MLNon-Hermitian,T_2019_NJP_QSLContinuousQuantumMeasurements,T_2020_PRX_StochasticTimeEvolution,T_2020_NatPhy_Time–informationUncertaintyRelations,E_2021_SciAdv_CrossoverQSL,T_2021_PRL_QSLChangingRate,T_2022_PRL_NewBound,T_2022_PRX_MTLimitsObservables,T_2022_PRXQauntum_SpeedLimitsMacroscopicTransitions,Hasegawa2023,T_2023_PRL_TopologicalSpeedLimit} 
as well as practical applications~\cite{T_2009_PRL_OptimalControlQSL, T_2015_PRL_QSLLeakageDecoherence,T_2016_SciRep_SpeedLimitsQuantumness,T_2017_PRL_Trade-offSpeedCostQSL,T_2017_PRL_ChangingPowerQB,T_2019_PRX_SpeedLimitStatePreparation,T_2020_PRR_AdiabaticQuantumComputation,T_2020_QSLInformationProduction,T_2022_NJP_ResourceSpeedLimits,T_2022_NJP_QSLQuantumGates,T_2022_NJP_QSLQuantumGravity,Howard2023,T_2023_PRX_ThermodynamicUnificationOptimalTransport} have since ensued. For isolated systems with a bounded energy spectrum (maximum spectral energy $E_{\rm max}$), a time-reversal symmetric version of the ML bound (dubbed $\rm ML^{\star}$) has also been characterized~\cite{T_2022_PRL_NewBound}, $t_\perp\geq t_{\rm ML^{\star}}\equiv \pi\hbar/[2(E_{\rm max}-E)]$. Their combination, results in a tight~\cite{Levitin_Toffoli_2009} (for pure states), \textit{unified}~\cite{T_2022_PRL_NewBound} minimal orthogonalization time $t_\perp\geq t_{\rm U}=\max\left[t_{\rm MT}, t_{\rm ML}, t_{\rm ML^{\star}}\right]$.

\begin{figure*}[t!]
    \begin{center}
    \includegraphics[width=0.85\textwidth]{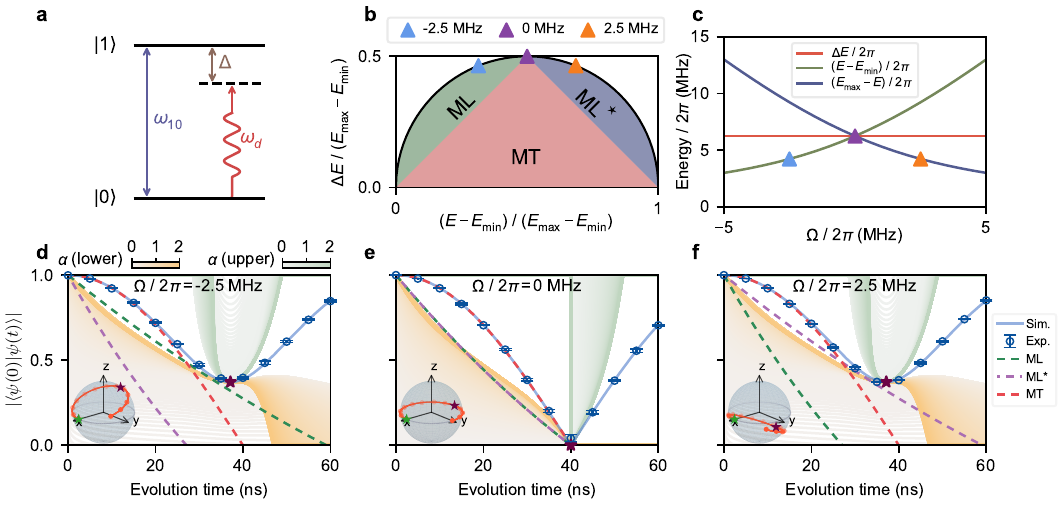}
    \end{center}
    \caption{\textbf{Quantum speed limit in a qubit system.}
    {\bf a}, Energy level representation of a driven qubit. Here, $\omega_{10}$ is the transition frequency between $|0\rangle$ and $|1\rangle$, $\omega_d$ is the frequency of microwave drive, and $\Delta$ is the detuning between $\omega_{10}$ and $\omega_d$. {\bf b}, QSL phase diagram in terms of the first and second moments of the energy, in the case of bounded energy spectra $\{E_n\}$, where the Bathia-Davis inequality~\cite{Bhatia_Davis_inequality_2000} applies. The different regions, MT, ML, and ML$^\star$ are further defined by whether the mean energy $E$ compares to $(E_{\rm min}+E_{\rm max})/2$. The blue, purple, and orange markers correspond to $\Omega/2\pi=-2.5$ MHz, 0 MHz, and 2.5 MHz, which are used in the experiment. {\bf c}, By choosing the initial state as $|\psi(0)\rangle=\frac{1}{\sqrt{2}}(|0\rangle+|1\rangle)$ we contrast the $\Omega$-dependency of $\Delta E$, $E-E_{\rm min}$ and $E_{\rm max}-E$; the smaller quantity at a given $\Omega$ defines the QSL region the dynamics is most tightly bounded at later times. {\bf d}-{\bf f}, Dynamics of overlap for $\Omega/2\pi=-2.5$ MHz, 0 MHz and 2.5 MHz. The blue circles (lines) are the experimental (theoretical) results of $|\langle\psi(0)|\psi(t)\rangle|$; dashed lines indicate the ML, $\rm ML^{\star}$ and MT bounds. The yellow and green shaded lines mapped by the color bars are the generalized lower and upper ML bounds. The green and purple stars highlight the initial state and the local minima of the theoretical $|\langle\psi(0)|\psi(t)\rangle|$, respectively. Corresponding dynamics trajectories are plotted on Bloch spheres in the lower left corner. Error bars stem from 5 repetitions of measurements.}
    \label{fig:qubit}
\end{figure*}

Experimental observation of these bounds has been put forward in single-particle systems, including a single atom on an optical trap~\cite{E_2021_SciAdv_CrossoverQSL} and a single transmon qubit in a cavity resonator~\cite{Wu_2024_testing_unified_bounds}. However, further experimental exploration of their generality and corresponding crossover in generic many-body systems remains uncharted territory. Likewise, also not often explored is the \textit{minimal} speed that the state overlap can change, which is crucial for quantum information processing. That is, the quantum dynamics of generic systems proceed, obeying maximal \textit{and} minimal speeds. Fortunately, remarkable progress in simulating many-body Hamiltonians using superconducting quantum circuits~\cite{E_2021_Science_InformationScrambling, E_2021_NatPhy_MBLSuperconductingQubits, Zhu2022PRL}, accompanied by the flexibility of initial state preparations has poised us to meet this demand. 

\begin{figure*}[t!]
    \begin{center}
    \includegraphics[width=0.85\textwidth]{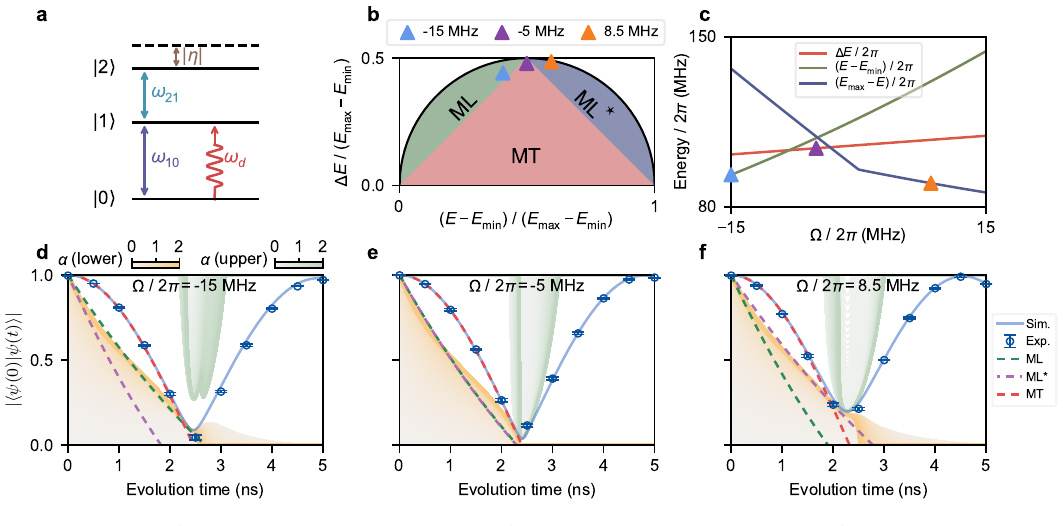}
    \end{center}
    \caption{{\bf Quantum speed limit in a qutrit system.} 
    {\bf a}, Energy level representation of a driven qutrit. The qutrit's nonlinearity is  $\eta/2\pi =-212$ MHz, and the transition frequencies satisfy the relation $\omega_{21}=2\omega_{10}+\eta$. Unlike the qubit case in Fig.~\ref{fig:qubit}, the qutrit is driven by a resonant microwave with $\omega_{d}=\omega_{10}$. 
    \textbf{b}, QSL phase diagram with markers annotated based on the values of $\Omega$. \textbf{c}, The $\Omega$-dependency of the quantities establishing the shortest orthogonalization time. 
    {\bf d}-{\bf f}, Dynamics of overlap for $\Omega/2\pi=-15$ MHz, -5 MHz and 8.5 MHz. Notations for the markers and lines are the same as the qubit case in Fig.~\ref{fig:qubit}. Error bars stem from 5 repetitions of measurements. Here, the initial state is $|\psi(0)\rangle=\frac{1}{\sqrt{10}}[|0\rangle+\frac{3}{\sqrt{2}}(|1\rangle+|2\rangle)]$. See Supplementary Section 7 for experimental data of longer time dynamics and Supplementary Section 11 for the effects of imperfect pulse.}
    \label{fig:qutrit}
\end{figure*}

Here, we further leverage the tunability of the quantum circuits to examine the crossover among different bounds. Crucially, our investigation relies on noticing that a \textit{single} Hamiltonian parameter can effectively govern which bound provides the tighter dynamics, even if fixing the initial state. Using this rationale, our study spans small (e.g., one qubit or one qutrit, Figs.~\ref{fig:qubit}\textbf{a} and \ref{fig:qutrit}\textbf{a}) and multi-qubit systems (such as one- and two-dimensional lattices shown in Figs.~\ref{fig:chain}\textbf{a} and ~\ref{fig:lattice}\textbf{a}), for which many-body effects are preponderant. 

\section*{Quantum speed limits in single qubit or qutrit systems}  \label{sec:qubits_qutrits}
For a general time-independent Hamiltonian with a bounded energy spectrum, $E_{\rm min} \leq \{E_n\} \leq E_{\rm max}$, the QSL is bounded by a unified limit  $t_{\rm U}=\max\left[\frac{\pi \hbar}{2\Delta E}, \frac{\pi \hbar}{2(E-E_{\rm min})}, \frac{\pi\hbar}{2(E_{\rm max} - E)}\right]$  \cite{T_2022_PRL_NewBound}. As a result of the Bhatia–Davis inequality~\cite{Bhatia_Davis_inequality_2000}, its energy uncertainty $\Delta E$ and mean energy $E$ satisfy the relation $\Delta E \leq \sqrt{(E_{\rm max} - E)(E-E_{\rm min})}$, which further establishes three dynamical regimes for the time evolution of any accessible initial state $|\psi(0)\rangle$. As shown in Fig.~\ref{fig:qubit}\textbf{b}, if $\Delta E < \min(E-E_{\rm min}, E_{\rm max}-E)$ the dynamics is dominated by the MT bound at all times; otherwise, i.e., $\Delta E > \min(E-E_{\rm min}, E_{\rm max}-E)$, a second condition, $E < (E_{\rm min} + E_{\rm max})/2$ [$E > (E_{\rm min} + E_{\rm max})/2$], emerges in which $t_{\rm ML}$ ($t_{\rm ML^*}$) limits the shortest orthogonalization time scale (see Supplementary Section 2 for further discussion).

To experimentally observe three dynamical regimes and the possible crossover between different bounds during the dynamics, we consider a qubit (qutrit) system under a microwave drive $\hat{H}_d=\Omega\cos(\omega_d t)$, where $\Omega$ is the drive amplitude and $\omega_d$ is the drive frequency. In this scenario, we can conveniently select which bound limits the speed of the system's dynamics by tuning a single Hamiltonian parameter $\Omega$ (see Supplementary Section 5 for experimental calibrations of $\Omega$), even if the initial state remains unchanged. In the rotating frame of drive frequency $\omega_d$, the Hamiltonian of a driven qubit is generically described by
\begin{equation}
    \hat{H}_{\rm qubit}/\hbar=\left( \begin{array}{cc} 0 & \Omega \\
    \Omega & \Delta \end{array} \right)\ ,
\label{eq:H_qubit}
\end{equation}
with $\Delta = \omega_{10}-\omega_d$ being the detuning between the qubit transition frequency $\omega_{10}$ and the drive frequency $\omega_d$ (Fig.~\ref{fig:qubit}\textbf{a}). By setting the initial state as $|\psi(0)\rangle=(|0\rangle+|1\rangle)/\sqrt{2}$, we show in Fig.~\ref{fig:qubit}\textbf{c} the $\Omega$-dependence of the three key quantities that classify which bound regime the dynamics belongs to: $\Delta E$, $E-E_{\rm min}$, and $E_{\rm max} - E$; the smallest of them defines the minimal orthogonalization time and establishes a constraint in the dynamics before $t_{\rm U}$ is reached. In experiments, we observe the dynamics of the overlap between the initial and time-evolved states, $\mathcal{F}\equiv|\avg{\psi(0)|\psi(t)}|$, which proceeds along the geodesic distance in the complex Hilbert space, $l_{\rm geo} = \arccos(\mathcal{F})$~\cite{T_1981_PRD_Distance,E_2021_SciAdv_CrossoverQSL}. It is bounded by ~\cite{Fleming1973, Bhattacharyya1983, T_2022_PRL_NewBound}
\begin{align}
   &\arccos(\mathcal{F}) \geq \notag \\ &\max\left(\frac{\Delta E t}{\hbar}, \sqrt{\frac{\pi (E-E_{\rm min})t}{2\hbar}}, \sqrt{\frac{\pi(E_{\rm max}-E)t}{2\hbar}}\right)\ .
   \label{eq:bounds_in_F}
\end{align}
By setting $\Omega/2\pi=-2.5$ MHz, 0 MHz, and 2.5 MHz, we show in Fig.~\ref{fig:qubit} \textbf{d}, \textbf{e}, and \textbf{f} how the actual dynamics under the Hamiltonian $\hat{H}_{\rm qubit}$ is tightly bounded by each of the constraints set above, at times $t \leq t_{\rm U}$, stemming from either the MT, ML or ML$^\star$ bounds -- at short times, however, MT is always the prevalent bound, see Supplementary Section 1.

In particular, the perfect orthogonalization 
for $\Omega=0$ MHz can be understood via the expectation values of the corresponding spin-1/2 Pauli matrices $\{\hat \sigma_\alpha\}$ ($\alpha=x,y,z$). In this case, $\hat H_{\rm qubit}$ is diagonal and thus proportional to $\hat \sigma_z$; the initial state $|\psi(0)\rangle = |+\rangle$, an eigenstate of $\hat \sigma_x$, evolves as described by the precession in the Bloch sphere around the $z$ quantization axis (inset in Fig.~\ref{fig:qubit}\textbf{e}). At $t=t_{\rm U}\simeq 40$ ns, $|\psi(t)\rangle = |-\rangle$, the second eigenstate of $\hat \sigma_x$, leading to ${\cal F} = 0$ (Fig.~\ref{fig:qubit}\textbf{e}). For $\Omega\neq0$ MHz, the corresponding representation of $\hat H_{\rm qubit}$ in terms of $\{\hat \sigma_{\alpha}\}$ results in an effective spin Hamiltonian with generic precession axis, precluding orthogonalization of the state, as seen in Fig.~\ref{fig:qubit}\textbf{d} and \textbf{f}.

The next level of complexity one can perform a similar analysis is on a three-level Hamiltonian, a qutrit. A transmon-type qubit is a multi-level system with a weak anharmonicity, and thus, a higher-energy state $|2\rangle$ will be involved in the system dynamics under the large microwave drive, which yields a driven qutrit Hamiltonian,
\begin{equation}
    \hat{H}_{\rm qutrit}/\hbar=\left( \begin{array}{ccc} 0 & \Omega & 0 \\
\Omega & 0 & \sqrt{2}\Omega \\
0 & \sqrt{2}\Omega & \eta\end{array} \right),
\label{eq:H_qutrit}
\end{equation}
where $\eta=\omega_{10} - \omega_{21}$, is the nonlinearity -- for our device, $\eta/2\pi\approx -212~{\rm MHz}$. Similarly, adjusting the drive amplitude allows the exploration of regimes in which the time evolution falls into the different QSL regions (Fig.~\ref{fig:qutrit}\textbf{b}), by checking the smallest of $\{\Delta E, E-E_{\rm min}, E_{\rm max} - E\}$ (Fig.~\ref{fig:qutrit}\textbf{c}). Here, the initial state is prepared as a superposition of the natural qutrit levels, $|\psi(0)\rangle=\frac{1}{\sqrt{10}}[|0\rangle+\frac{3}{\sqrt{2}}(|1\rangle+|2\rangle)]$ (see Supplementary Section 6 for details of qutrit gate calibrations).

Albeit remarkably restricting, the bounds in Eq.~\eqref{eq:bounds_in_F} are not always the tightest in actual dynamics. Generalizations of the ML (ML$^{*}$) bound~\cite{T_2022_PRL_NewBound}, based on a generic L$^\alpha$-norm, $E_\alpha \equiv \langle (\hat H - E_{\rm min})^\alpha\rangle^{1/\alpha}$ $\left(E^*_\alpha \equiv \langle (E_{\rm max} - \hat H)^\alpha\rangle^{1/\alpha}\right)$ with $\alpha \in \mathbb{R}^+$, have been demonstrated~\cite{T_2005_LMP_GeneralizedML, T_2006_PRA_GeneralizedML}, resulting in orthogonalization times $t_{{\rm ML},\alpha}\equiv \pi\hbar/(2^{1/\alpha}E_\alpha)$ 
$\left(t_{{\rm ML}^*,\alpha}\equiv \pi\hbar/(2^{1/\alpha}E_\alpha^*)]\right)$. A simple derivation in the Supplementary Section 3 shows that, in general, the shortest orthogonalization time satisfies,
\begin{equation}
    t_\perp \geq \left\{\gamma_\alpha({\cal F}, \theta^{\prime})^{\frac{1}{\alpha}}\frac{\pi\hbar}{2^{1/\alpha}E_\alpha}, \gamma_\alpha({\cal F}, \theta^{\prime\prime})^{\frac{1}{\alpha}}\frac{\pi\hbar}{2^{1/\alpha}E_\alpha^*}\right\}\ ,
\end{equation}
where the prefactor $\gamma_\alpha({\cal F}, \theta^{\prime}) \equiv \max\{0, 1-{\cal F}d^{\prime}_{\alpha}\}$ [$\gamma_\alpha({\cal F}, \theta^{\prime\prime}) \equiv \max\{0, 1-{\cal F}d^{\prime\prime}_{\alpha}\}$]. Here we define $d^{\prime}_{\alpha} \equiv \cos\theta^{\prime}-\frac{2\alpha}{\pi}\sin\theta^{\prime}$ ($d^{\prime\prime}_{\alpha} \equiv \cos\theta^{\prime\prime}-\frac{2\alpha}{\pi}\sin\theta^{\prime\prime}$). $\theta^{\prime}=\theta+E_{\rm min}t\ $  and $\theta^{\prime\prime}=-\theta-E_{\max}t$ are defined by the time-dependent phase that accumulates in the overlap: $\langle \psi(0)|\psi(t)\rangle = {\cal F}e^{{\rm i}\theta}$.
Similar to the more known bounds [Eq.~\eqref{eq:bounds_in_F}], the overlap itself is always bounded depending on the sign of $d^{\prime}_{\alpha}$ 
:
\begin{subequations}
\begin{align}
1 &\geq \mathcal{F} \geq B_{\rm low}^{\prime,\alpha},\ d^{\prime}_{\alpha}>0, \\
B_{\rm up}^{\prime,\alpha} &\geq \mathcal{F} \geq 0,\ d^{\prime}_{\alpha}<0, 
\end{align}
\label{eq:gen_lower_bound1}
\end{subequations}
where $B_{\rm low}^{\prime,\alpha}\equiv\max[0,\frac{1-\frac{2t^{\alpha}}{(\pi\hbar)^{\alpha}}\langle(\hat{H}-E_{\rm min})^{\alpha}\rangle}{d^{\prime}_{\alpha}}]$ for $d^{\prime}_{\alpha}>0$ [$B_{\rm up}^{\prime,\alpha}\equiv\min[1,\frac{1-\frac{2t^{\alpha}}{(\pi\hbar)^{\alpha}}\langle(\hat{H}-E_{\rm min})^{\alpha}\rangle}{d^{\prime}_{\alpha}}]$ for $d^{\prime}_{\alpha}<0$]. Similarly, based on the sign of $d^{\prime\prime}_{\alpha}$ one can have another set of bounds:
\begin{subequations}
\begin{align}
1 &\geq \mathcal{F} \geq B_{\rm low}^{\prime\prime,\alpha},\ d^{\prime\prime}_{\alpha}>0, \\
B_{\rm up}^{\prime\prime,\alpha} &\geq \mathcal{F} \geq 0,\ d^{\prime\prime}_{\alpha}<0, 
\end{align}
\label{eq:gen_lower_bound2}
\end{subequations}
where $B_{\rm low}^{\prime\prime,\alpha}\equiv\max[0,\frac{1-\frac{2t^{\alpha}}{(\pi\hbar)^{\alpha}}\langle(E_{\rm max}-\hat{H})^{\alpha}\rangle}{d^{\prime\prime}_{\alpha}}]$ for $d^{\prime\prime}_{\alpha}>0$ [$B_{\rm up}^{\prime\prime,\alpha}\equiv\min[1,\frac{1-\frac{2t^{\alpha}}{(\pi\hbar)^{\alpha}}\langle(E_{\rm max}-\hat{H})^{\alpha}\rangle}{d^{\prime\prime}_{\alpha}}]$ for $d^{\prime\prime}_{\alpha}<0$]. Combining Eq.~(\ref{eq:gen_lower_bound1}) and (\ref{eq:gen_lower_bound2}) gives the final bound of $\cal{F}$ for a specific $\alpha$, summarized in Table~\ref{tab:GML}. By calculating the bounds at different $\alpha$, one can get a set of lower and upper bounds for $\cal{F}$, which are depicted together with the MT, ML and ML$^{\star}$ bounds in Figs.~\ref{fig:qubit} - \ref{fig:lattice}.
\begin{table}[htb]
    \centering
    \caption{How the overlap $\cal{F}$ is bounded by the generalized lower and upper bounds for $\alpha$, and the sign of $d^{\prime}_{\alpha}$ and $d^{\prime\prime}_{\alpha}$.}
    \begin{tabular}{|c|c|c|}
        \hline
         \diagbox{$d^{\prime}_{\alpha}$}{$d^{\prime\prime}_{\alpha}$} &  +  & -  \\ \hline
          + &  $1\geq\mathcal{F}\geq\max[B_{\rm low}^{\prime,\alpha},B_{\rm low}^{\prime\prime, \alpha}]$  &  $B_{\rm up}^{\prime\prime, \alpha}\geq\mathcal{F}\geq B_{\rm low}^{\prime, \alpha}$  \\ \hline
          - &  $B_{\rm up}^{\prime, \alpha}\geq\mathcal{F}\geq B_{\rm low}^{\prime\prime, \alpha}$  &  $\min[B_{\rm up}^{\prime, \alpha},B_{\rm up}^{\prime\prime, \alpha}]\geq\mathcal{F}\geq0$   \\ \hline
    \end{tabular}
    \label{tab:GML}
\end{table}

In Fig.~\ref{fig:qubit}\textbf{d} - \textbf{f} and Fig.~\ref{fig:qutrit}\textbf{d} - \textbf{f}, we show how the generalized ML bounds in {Table~\ref{tab:GML}} limit the speed of overlap dynamics. First, the generalized bounds typically exhibit tighter lower bounds in overlap, particularly at large times. Second, the upper bound in the dynamics coming from {Table~\ref{tab:GML}} makes the instants where the smallest overlap is attainable unique: the dynamics have no room but to follow a tight `speed,' bounded from above \textit{and} below. Moreover, these tight regions can periodically occur in time, a direct consequence of the periodic nature of the precessing dynamics in the (generalized) Bloch sphere (see Supplementary Section 7).

\section*{Quantum speed limits in many-body systems}
\begin{figure}[t!]
    \begin{center}
    \includegraphics[width=1.0\columnwidth]{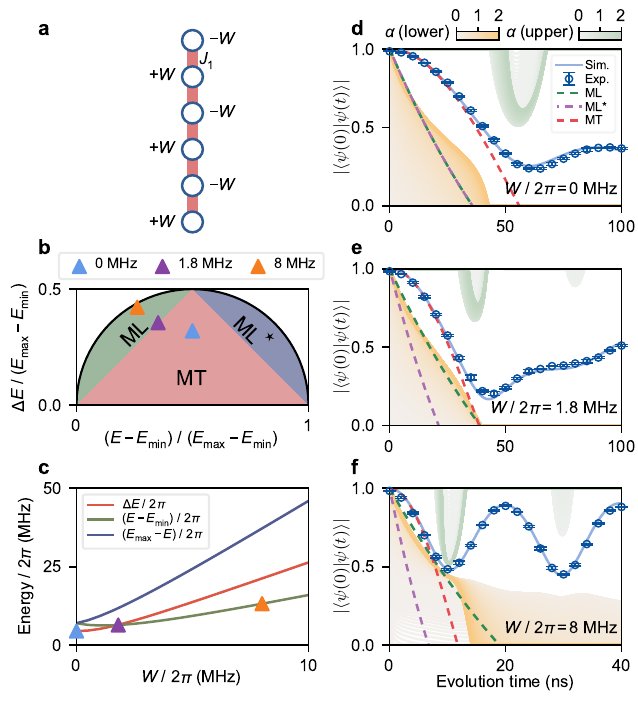}
    \end{center}
    \caption{{\bf Quantum speed limit for the 1d qubit chain.} {\bf a}, Schematic of the 1d qubit chain. Each circle represents a qubit with onsite potential $\pm W$. Neighbor qubits are coupled with coupling strengths $J_1/2\pi=-2$ MHz, denoted by pink lines. \textbf{b}, QSL phase diagram. \textbf{c}, $W$-dependency of the quantities establishes the shortest orthogonalization time bound. The initial state is chosen as the unbalanced superposition of two density wave product states $|\psi(0)\rangle=\frac{1}{2}(|101010\rangle+\sqrt{3}|010101\rangle)$. 
    {\bf d}-{\bf f}, Dynamics of overlap for $W/2\pi=0$ MHz, 1.8 MHz and 8 MHz, respectively. Notation for markers and lines is similar to Figs.~\ref{fig:qubit} and \ref{fig:qutrit}. Error bars stem from 5 repetitions of measurements.}
    \label{fig:chain}
\end{figure}

Nothing so far is unique to simple quantum mechanical systems. The only constraint is that the Hamiltonian governing the dynamics has a bounded spectrum so that ML, ML$^*$ bounds and their generalizations are well-defined. As a result, the constraints we derive should also apply to inherently many-body Hamiltonians, which can similarly be emulated using a network of superconducting qubits but have not been experimentally explored on what concerns the speed of evolution of its many-body states in the Hilbert space.

To that end, leveraging the inter-qubit coupling tunability, their onsite energies, and the flexibility in preparing multi-qubit entangled states (see Supplementary Section 4 for more device information), we emulate the quantum XY-model in a one dimensional (1d) chain of six qubits:
\begin{equation}
    \hat{H}_{\rm 1d}/\hbar=J_1\sum_{\avg{ij}}(\hat{\sigma}_i^+\hat{\sigma}_j^-+\hat{\sigma}_i^-\hat{\sigma}_j^+)+\sum_iW_i\hat{\sigma}_i^+\hat{\sigma}_i^-\ ,
\label{eq:H_chain} 
\end{equation}
where $J_1/2\pi=-2$ MHz (see Supplementary Section 8) gives the hopping energy scale for excitations of nearest-neighbor qubits; $W_i$ is the on-site potential of the $i$-th qubit. In this case, we take $W_i = (-1)^i W$, forming a staggered potential pattern (Fig.~\ref{fig:chain}\textbf{a}) and the initial state is prepared as an entangled state $|\psi(0)\rangle=\frac{1}{2}(|101010\rangle+\sqrt{3}|010101\rangle)$ (see Supplementary Section 9 for the state preparation circuit). The single Hamiltonian parameter controlling the restriction on the dynamics given by different bounds is the amplitude $W$. In Fig.~\ref{fig:chain}\textbf{d} - \textbf{f}, we show that the dominant bound can be changed by adjusting $W/2\pi$ from 0 MHz to 8 MHz (see Supplementary Section 10
for the measurement of overlap).

We make a closer inspection by selecting three representative values, $W/2\pi = 0$ MHz, 1.8 MHz, and 8 MHz, regimes into which the most restrictive bound changes from MT to the ML one (see Fig.~\ref{fig:chain}\textbf{c} for the corresponding values of $\Delta E$, $E - E_{\rm min}$ and $E_{\rm max}-E$ vs.~$W$). For $W/2\pi = 0$ MHz or 1.8 MHz, Fig.~\ref{fig:chain}\textbf{d} and \textbf{e}, the lower generalized bounds are less restrictive than the original MT bound. However, as we delve deeper into the ML regime (Fig.~\ref{fig:chain}\textbf{f}), the situation changes. Beyond a crossing time $t_c \approx $ 7 ns, ML exhibits the tightest lower bound, being superseded by its generalized version at longer times. Notably, at the times of lowest overlap, the upper generalized bound constrains $\cal F$ such that its values are tightly limited from above \textit{and} below, as for the qubit and qutrit Hamiltonians.

\begin{figure*}[t!]
    \begin{center}
    \includegraphics[width=0.85\textwidth]{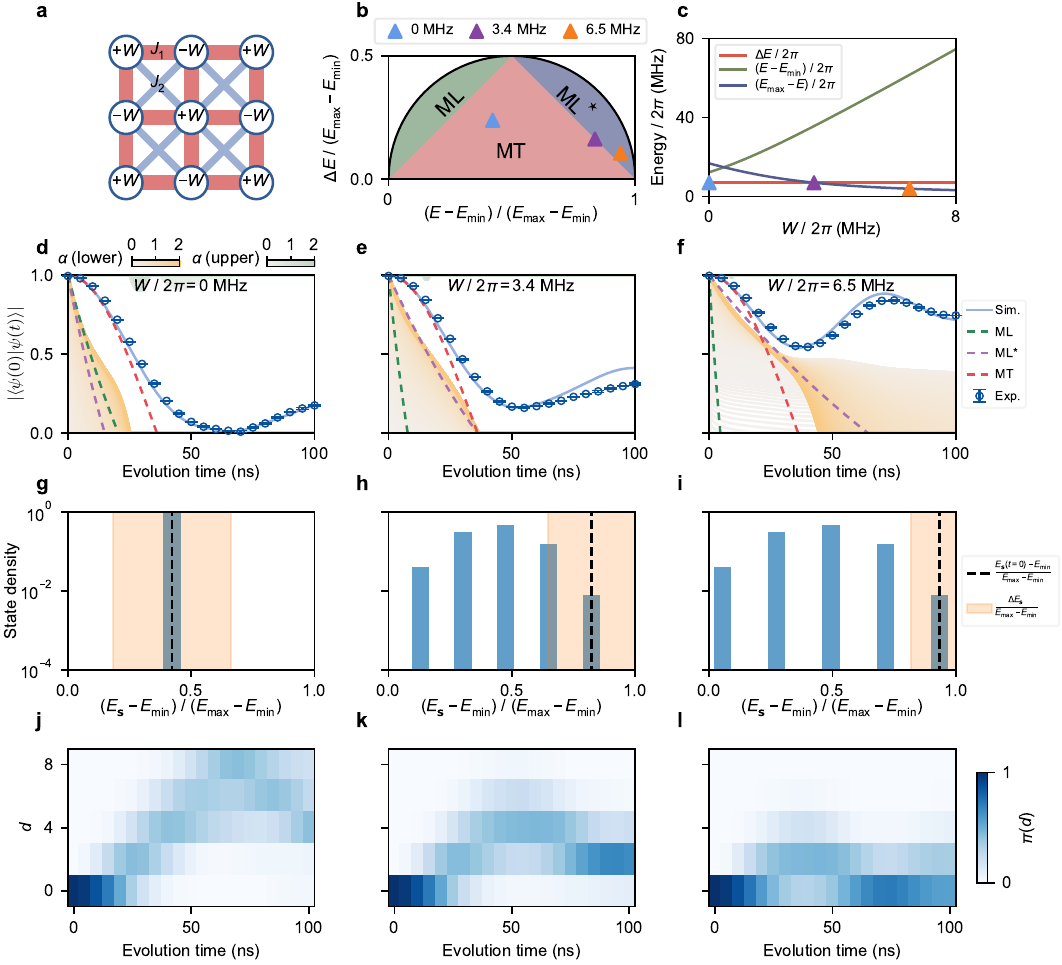}
    \end{center}
    \caption{{\bf Quantum speed limit for 2d square qubit lattice.} 
    {\bf a}, 9-qubit ($3\times3$) lattice with (next-) nearest neighbor coupling $J_1/2\pi = -2$ MHz ($J_2/2\pi=0.597$ MHz) in the presence of a checkerboard potential with amplitude $W$, which is tuned to establish the crossover from MT to an ML$^*$-dominated bound. 
    \textbf{b}, QSL phase diagram. 
    \textbf{c}, $W$-dependency of the quantities establishing which is the shortest orthogonalization time. Here, the initial state is a single product state featuring five excitations initialized in the qubits with $+W$ energy. 
    {\bf d}-{\bf f}, Dynamics of overlap for $W/2\pi=0$ MHz, 3.4 MHz and 6.5 MHz; here notations for markers and lines are similar to Figs.~\ref{fig:qubit} and \ref{fig:qutrit} and error bars stem from 5 repetitions of measurements.  {\bf g}-{\bf i}, Density of Fock states in the rescaled energy spectrum, highlighting the mean energy $E=E_{\mathbf{s}}$ (vertical dashed line) and the energy uncertainty $\Delta E$ (shaded orange region). {\bf j}-{\bf l}, Dynamics in Fock space, in terms of the wave-packet probability at the Hamming distance $d$.}
    \label{fig:lattice}
\end{figure*}

Additionally, this type of integrable Hamiltonian has been extensively investigated~\cite{Cazalilla2011} since one can recast it in terms of free fermions via the Jordan-Wigner transformation~\cite{JordanWigner1928}; as a result, revivals of overlap are expected. While the generalized orthogonality times are limited and typically shorter than the ones provided by the standard bounds, they are significantly constraining when deep into the ML regime -- the upper generalized bounds also exhibit revivals closely following the oscillations in the overlap (see Supplementary Section 13 for more details).

Departing from integrability, we next emulate a similar Hamiltonian, using a two-dimensional (2d) structure featuring nine qubits. The Hamiltonian, schematically represented in Fig.~\ref{fig:lattice}\textbf{a}, reads
\begin{equation}
\begin{aligned}
    \hat{H}_{\rm 2d}/\hbar&=J_1\sum_{\avg{ij}}(\hat{\sigma}_i^+\hat{\sigma}_j^-+\hat{\sigma}_i^-\hat{\sigma}_j^+) \\ 
    &+J_2\sum_{\avg{\avg{ij}}}(\hat{\sigma}_i^+\hat{\sigma}_j^-+\hat{\sigma}_i^-\hat{\sigma}_j^+)+\sum_iW_i\hat{\sigma}_i^+\hat{\sigma}_i^-,
\end{aligned}
\label{eq:H_lattice}
\end{equation}
where $\avg{\cdot}$ and $\avg{\avg{\cdot}}$ denote the nearest and next-nearest neighbor hopping qubit excitations with amplitudes $J_1$ and $J_2$, respectively; $W_i = (-1)^{i_x+i_y}W$ is the checkerboard on-site potential of the qubit $i$. This is a typical nonintegrable Hamiltonian, where thermalization takes place~\cite{Rigol2008} and the time evolution proceeds by exploring the Hilbert space diffusively. 

We show results for five excitations, described as an initial product state where the excitations are initialized in the qubits with onsite energy $+W$. As before, $W$ is the tuning parameter that, if increased, leads to a regime whose dynamics, constrained initially by the MT bound, become limited by the ML$^*$ one at a value $W/2\pi = 3.4$ MHz (Fig.~\ref{fig:lattice}\textbf{b}). It is easy to see that this is the case: The initial product state, which is not an eigenstate of Hamiltonian \eqref{eq:H_lattice}, has associated mean energy $E=\langle\psi(0)|\hat H_{\rm 2d}|\psi(0)\rangle = 5W$. For sufficiently large $W$, this state exhibits $E$ close to the maximal spectral energy of $\hat H_{\rm 2d}$, such that $t_{\rm ML^{\star}}\equiv \pi\hbar/[2(E_{\rm max}-E)]$ is large.

The overlap dynamics for representative values of $W$ (Fig.~\ref{fig:lattice}\textbf{d} - \textbf{f}) shows that bounds are typically less stringent, with the exception being when deep in the ML$^*$ regime, wherein generalized ML$^*$ bounds (Table~\ref{tab:GML}) lower-limit it, exhibiting tighter bounds than the traditional ML$^*$ case [Eq.~\eqref{eq:bounds_in_F}]. That the bounds are typically less constraining follows from the extensiveness in system size of the energy and its uncertainty $\Delta E$, i.e., for large system sizes preserving the density of excitations, one expects the orthogonalization times to become vanishingly small (see Supplementary Section 12). Yet, simultaneously, the rate of change of overlap increases in this limit such that the bounds at ${\cal F}>0$ are always relevant for the corresponding dynamics.

For finite system sizes, the crossover of the bounds can be characterized by inspecting how the contribution of the different Fock states $|\mathbf{s}\rangle$ to $|\psi(t)\rangle$ evolves with time and with the amplitude of the checkerboard potential $W$. The latter renders that the Fock state energies turn stratified, counting the number of excitations in the $\pm W$ qubits, $E_{\mathbf{s}} = \sum_i W_i \langle \hat \sigma_i^+ \hat \sigma_i^-\rangle$. At $W/2\pi=0$ MHz, all product states are degenerate with $E_{\mathbf{s}} = 0$ MHz (Fig.~\ref{fig:lattice}\textbf{g}); the energy uncertainty encompasses all of them, allowing the dynamics to explore all possible states. At $W/2\pi = 3.4$ MHz, wherein $t_{\rm MT} \simeq t_{\rm ML^*}$, $\Delta E$ is sufficient to encompass only the initial state and the Fock states that exhibit one single excitation in a $-W$ qubit and the remaining ones in the $+W$ qubits with $E_{\mathbf{s}} = 3W$, forming a 20-fold degenerate subspace. This conclusion, that $\Delta E \simeq 2W$ at the crossover point, carries over to larger system sizes, also in the one-dimensional case (see Supplementary Section 14). Finally, further increasing the checkerboard energy offset to $W/2\pi = 6.5$ MHz makes $E_{\mathbf{s}}$ to split further such that the energy uncertainty only encloses the initial state with $E_{\mathbf{s}} = 5W$.

Verification of Fock space exploration can be quantified by the $L_1$-norm distance (Hamming distance) from the initial state $|\psi(0)\rangle \equiv |\mathbf{s}_0\rangle$ to any other Fock state $|\mathbf{s}\rangle$: $D(\mathbf{s},\mathbf{s}_0)=\sum_{i}|s^{i}-s_{0}^{i}|$, where $s^{i}\equiv\langle \mathbf{s}|\hat \sigma_i^+\hat \sigma_i^-|\mathbf{s}\rangle = 0$ or 1, denotes the presence or absence of an excitation in qubit $i$. The many-body Fock space dynamics are identified by the probability of finding an excitation at a distance $D(\mathbf{s}, \mathbf{s}_0)\equiv d$  at an instant of time $t$ \cite{ET_2022_NatPhys_FockSpaceDynamics},
\begin{equation}
 \Pi(d, t) = \sum_{\mathbf{s}\in \{D(\mathbf{s}, \mathbf{s}_0)=d\}}|\langle\mathbf{s}|e^{-\frac{{\rm i}\hat H_{\rm 2d}t}{\hbar}}|{\mathbf{s}_0}\rangle|^2  \ .
\end{equation}
This is shown in Fig.~\ref{fig:lattice}\textbf{j} - \textbf{l}, which describes the propagation of the wave-packet over the whole Fock space within the MT-limited regime ($W/2\pi = 0$ MHz), whereas, in the ML$^{*}$ governed one, this exploration is significantly limited to the states with $d \lesssim 4$, i.e., approximately in the manifold $E_{\mathbf{s}}= 1W,\  3W$ and $5W$.

\section*{Discussion and Outlook \label{sec:conclusions}}
\indent Quantum speed limits have been long introduced~\cite{T_1945_JPUSSR_OriginalMTbound, T_1998_PhysicaD_MLbound}, and their importance relies on the bounds they provide to any operation performed on a quantum state. Applications of that range from how fast a quantum gate operates in a quantum computer~\cite{T_2002_PRL_ComputationalCapacityUniverse, Ashhab_2012_theory, Howard2023} to how to maximize the charging speed of a quantum battery~\cite{T_2017_PRL_ChangingPowerQB, Julia-Farre2020}. Besides the maximum rate of change typically investigated~\cite{E_2021_SciAdv_CrossoverQSL, Wu_2024_testing_unified_bounds}, we showed here the existence of (generalized) minimal speeds a quantum state needs to obey under unitary conditions -- a related idea has been investigated for the minimal discharge speed of a quantum battery~\cite{Mohan2021}. Consequently, in various isolated quantum systems, state space exploration proceeds at a restricted speed from above and below. In complex state spaces, like those of large many-body quantum systems, bounds can be less stringent and restricted to short time scales -- a direct outcome of the fact that orthogonalization is facilitated with a larger number of states to explore.

Our systematic experimental characterization in structured single or many-body problems establishes the stage for future analysis under dissipation, whether it is engineered or not~\cite{Harrington2022, Mi2024}. In particular, it also allows testing the cases in which an open quantum system described by a Markovian bath has dynamics at short times defined by an effective non-Hermitian Hamiltonian~\cite{Naghiloo2019, Yang2019} and the associated quantum speed limit derived for these cases~\cite{Impens2021, Hornedal2024}.  

\section*{ACKNOWLEDGMENTS \label{sec:acknowledgements}}
\indent The device was fabricated at the Micro-Nano Fabrication Center of Zhejiang University. 
The instruments used to control and readout qubits are developed by Chip-hop Ltd.
We acknowledge support from the National Natural Science Foundation of China (Grant Nos. 12274368, U2230402, 12174342, 12322414, 12274367, 12247101), the Innovation Program for Quantum Science and Technology (Grant
No. 2021ZD0300200), the National Key R\&D Program of China (Grant No. 2023YFB4502600), the Zhejiang Provincial Natural Science Foundation of China (Grant Nos. LR24A040002, LDQ23A040001), and Zhejiang Pioneer (Jianbing) Project (Grant No. 2023C01036). R.M.~acknowledges support from the T$_{\rm c}$SUH Welch Professorship Award; part of the calculations used resources from the Research Computing Data Core at the University of Houston.
\\
\appendix

\bibliography{main}

\end{document}


\title{Supplementary Information for\\
Quantum highway: Observation of minimal and maximal speed limits for few and many-body states}
\maketitle
\onecolumngrid

\tableofcontents
\beginsupplement


\section{Mandelstam-Tamm bound's prevalence at short times \label{app:short_time}}

In the main text, we argue that the minimum orthogonalization time can be easily adjusted by modifying a Hamiltonian parameter, thereby changing the relative amplitude of the mean energy $E$ and variance $\Delta E$. However, the overlap ${\cal F} = |\langle \psi(0)|\psi(t)\rangle|$ remains bounded even before orthogonalization occurs at time 
$t=t_{\rm U}$. Specifically, we demonstrate below that $\cal{F}$ is initially governed by the Mandelstam-Tamm (MT) bound, which is related to $\Delta E$. To see this, we first check the difference between the MT and ML bounds at short times [see Eq.~(2) in the main text],
\begin{align}
    \delta \equiv \cos\left(\frac{\Delta Et}{\hbar}\right)-\cos\left(\sqrt{\frac{\pi (E-E_{\rm min})t}{2\hbar}}\right)\ .
\end{align}
Making use of a simple trigonometric identity,
\begin{equation}
    \cos x-\cos y=-2\sin\left(\frac{x-y}{2}\right)\sin\left(\frac{x+y}{2}\right)\ ,
\end{equation}
we see that at $t\ll1$,
\begin{equation}
\begin{aligned}
  \delta&=-2\sin\left(\frac{\frac{\Delta Et}{\hbar}-\sqrt{\frac{\pi (E-E_{\rm min})t}{2\hbar}}}{2}\right)\sin\left(\frac{\frac{\Delta Et}{\hbar}+\sqrt{\frac{\pi (E-E_{\rm min})t}{2\hbar}}}{2}\right)\\
  &\simeq-2\sin\left(\frac{-\sqrt{\frac{\pi (E-E_{\rm min})t}{2\hbar}}}{2}\right)\sin\left(\frac{\sqrt{\frac{\pi (E-E_{\rm min})t}{2\hbar}}}{2}\right)\\
  &=2\sin^2\left(\frac{\sqrt{\frac{\pi (E-E_{\rm min})t}{2\hbar}}}{2}\right)\geq0\ ,
\end{aligned}
\label{eq:MTbeatsML}
\end{equation}
where we used that $\sqrt{t}\gg t$ for $t\ll1$. This shows that the MT bound is tighter, always dominating at short times. Consequently, this also explains the crossover between the MT and ML bounds when $E -E_{\rm min} < \Delta E$ (ML-bounded dynamics) -- see, for example, Fig.~1{\bf d} in the main text where $t_c \simeq 27$ ns. Likewise, one can extend this difference between MT and the ML$^{\star}$ bounds by a similar analysis. In summary, the MT bound always dominates against the other two within short time scales.

\section{Blockage of the MT bound \label{app:blockageMT}}
A second significant theoretical result that we revise here is that a special class of states can prevent the MT bound from being the tighter one at later times. For instance, given the normalized initial state $|\psi(0)\rangle=\sum_ic_i|E_i\rangle$, in which $\{|E_i\rangle\}$ is the energy eigenbasis of the Hamiltonian with the corresponding coefficient $c_i$, the mean energy and variance are $E=\sum_i|c_i|^2E_i$ and $\Delta E=\sqrt{\sum_i|c_i|^2\left(E_i^2-E^2\right)}$, respectively. If further assuming that only two different eigenstates contribute to the initial state, i.e., $|\psi(0)\rangle=c_m|E_m\rangle+c_n|E_n\rangle$, $(|E_m\rangle$ and $|E_n\rangle$ are arbitrary eigenstates), then we have the following theorem:
\vskip0.1in
\noindent\textbf{Theorem} \ \ Blockage of MT bound\\
\textit{The position of the initial states in the QSL phase diagram saturates the Bathia-Davis inequality~\cite{Bhatia_Davis_inequality_2000} if and only if the initial state is expanded only by the ground and the highest-energy state: $|\psi(0)\rangle=c_{\rm min}|E_{\rm min}\rangle+c_{\rm max}|E_{\max}\rangle$, where $E_{\rm min}$ and $E_{\rm max}$ are the ground and highest spectral energies respectively.}
\vskip0.1in
\noindent \textit{Proof:} Given $|\psi(0)\rangle=c_m|E_m\rangle+c_n|E_n\rangle$, the mean energy and variance are:
\begin{equation}
\begin{aligned}
    E&=|c_m|^2E_m+|c_n|^2E_n, \\
    \Delta E^2 &=|c_m|^2E^2_m+|c_n|^2E^2_n-E^2.
\end{aligned}
\end{equation}
If one further defines the difference $\mathcal{D}=(E_{\rm max}-E)(E-E_{\rm min})-\Delta E^2$, intuited by  Bathia-Davis inequality on variances, $\Delta E \leq \sqrt{(E_{\rm max} - E)(E-E_{\rm min})}$, which sets the upper boundary of the phase diagram, one has
\begin{equation}
    \mathcal{D}=|c_m|^2(E_m - E_{\rm min})(E_{\rm max}-E_m)+|c_n|^2(E_n - E_{\rm min})(E_{\rm max}-E_n).
\end{equation}
When $\mathcal{D}=0$, the Bathia-Davis' inequality is fulfilled. Thus leads to
\begin{equation}
    |c_m|^2r_m(r_m-1)=|c_n|^2r_n(1-r_n),
    \label{appeq:rm_rn}
\end{equation}
where $0\leq r_m=(E_m-E_{\rm min})/(E_{\rm max}-E_{\rm min})\leq1$ and $0\leq r_n=(E_n-E_{\rm min})/(E_{\rm max}-E_{\rm min})\leq1$. Equation \eqref{appeq:rm_rn} can only be satisfied when $r_m$ and $r_n$ are either 0 or 1, which implies that $|E_m\rangle$ and $|E_n\rangle$ are either $|E_{\rm min}\rangle$ or $|E_{\max}\rangle$. As a pedagogical case, one can easily check the position of every pure initial state of a two-level system (Fig.~1\textbf{b}) satisfies Bathia-Davis' inequality and lies on the curving boundary of the phase diagram.  

\section{Generalization of the ML bound \label{app:hign-order-bound}}
\indent Ref.~\cite{T_2006_PRA_GeneralizedML} shows a generalized ML bound for $t_{\perp}$, which reads
\begin{equation}
    t_{\perp} \geq \frac{\pi\hbar}{2^{1/\alpha}\langle(\hat{H}-E_{\rm min})^{\alpha}\rangle^{1/\alpha}},~\alpha>0.
    \label{appeq:high-order-bound}
\end{equation}
Based on this, we can derive another bound similar to Eq.~\eqref{appeq:high-order-bound} except that there is an additional prefactor valid for all $t$. Given the Hamiltonian $\hat{H}$ and its eigenstates ${|E_n\rangle}$, initial and time-evolved states are expanded as $|\psi(0)\rangle=\sum_n c_n|E_n\rangle$ and $|\psi(t)\rangle=\sum_n c_ne^{-{\rm i}E_nt}|E_n\rangle$ ($\hbar=1$). Following a similar procedure as in the Appendix of Ref.~\cite{T_2003_PRA_EntanglementSpeedupQSL}, we define 
\begin{equation}
    \langle\psi(0)|\psi(t)\rangle=\mathcal{F} e^{{\rm i}\theta},
    \label{appeq:overlap}
\end{equation}
where $\mathcal{F}=|\langle\psi(0)|\psi(t)\rangle|$ is the overlap defined in the main text. For simplicity, we can promote an energy shift, such that one defines a semi-positive definite Hamiltonian $\hat{H}^{\prime} = \hat{H} - E_{\min}$ ($E_{\min}$ is the ground state energy of $\hat{H}$) and the corresponding overlap
\begin{equation}
    \langle\psi(0)|\psi^{\prime}(t)\rangle=\mathcal{F} e^{{\rm i}\theta^{\prime}},
    \label{appeq:overlap_prime}
\end{equation}
where $|\psi^{\prime}(t)\rangle=\sum_nc_ne^{-{\rm i}(E_n-E_{\min})t}|E_n\rangle$ -- we can see that $\theta^{\prime}=\theta+E_{\min}t$. Additionally, we can further have 
\begin{equation}
    \begin{aligned}
        \sum_n|c_n|^2\cos[(E_n-E_{\min})t] &= \mathcal{F} \cos\theta^{\prime},\\
        \sum_n|c_n|^2\sin[(E_n-E_{\min})t] &= -\mathcal{F}\sin\theta^{\prime}.
        \label{appeq:inequality0}
    \end{aligned}
\end{equation}
Now, borrowing the trigonometric inequality~\cite{T_2006_PRA_GeneralizedML}
\begin{equation}
    x^{\alpha}-\frac{\pi^{\alpha}}{2}+\frac{\pi^{\alpha}}{2}\cos x + \alpha\pi^{\alpha-1}\sin x\geq0,    
\end{equation}
which is valid for $x\geq0,\alpha>0$, we replace $x$ with $(E_n-E_{\min})t$ to obtain 
\begin{equation}
\begin{aligned}
    &\sum_n|c_n|^2\left(\cos[(E_n-E_{\min})t] + \frac{2\alpha}{\pi}\sin[(E_n-E_{\min})t]\right)\\
    &\geq1-\frac{2t^{\alpha}}{\pi^{\alpha}}\sum_n|c_n|^2(E_n-E_{\min})^{\alpha}\\
    &=1-\frac{2t^{\alpha}}{\pi^{\alpha}}\langle(\hat{H}-E_{\min})^{\alpha}\rangle.
    \label{appeq:inequality1}
\end{aligned}
\end{equation}
Combining Eqs.~\eqref{appeq:inequality0} and \eqref{appeq:inequality1}, we obtain 
\begin{equation}
    \mathcal{F}(\cos\theta^{\prime}-\frac{2\alpha}{\pi}\sin\theta^{\prime})\geq 1-\frac{2t^{\alpha}}{\pi^{\alpha}}\langle(\hat{H}-E_{\min})^{\alpha}\rangle\ ,
    \label{appeq:fidelity1}
\end{equation}
leading to 
\begin{equation}
    t^{\alpha} \geq [1-\mathcal{F}(\cos\theta^{\prime}-\frac{2\alpha}{\pi}\sin\theta^{\prime})]\frac{\pi^{\alpha}}{2\langle(\hat{H}-E_{\min})^{\alpha}\rangle}.
\end{equation}
As a result, any time $t$ satisfies,
\begin{equation}
    t \geq \gamma(\mathcal{F}, \theta^{\prime})^{1/\alpha}\frac{\pi}{2^{1/\alpha}\langle(\hat{H}-E_{\min})^{\alpha}\rangle^{1/\alpha}}\ , 
\end{equation}
where $\gamma(\mathcal{F}, \theta^{\prime}) \equiv \max[0, 1-\mathcal{F}(\cos\theta^{\prime}-\frac{2\alpha}{\pi}\sin\theta^{\prime})]$.
 In particular, these equations lead to (generalized) overlap bounds even before orthogonalization occurs. When $\cos\theta^{\prime}-\frac{2\alpha}{\pi}\sin\theta^{\prime}>0$, we can have, based on Eq.~(\ref{appeq:fidelity1}), a lower bound for the overlap $\mathcal{F}$
\begin{equation}
    \mathcal{F} \geq \frac{1-\frac{2t^{\alpha}}{\pi^{\alpha}}\langle(\hat{H}-E_{\min})^{\alpha}\rangle}{\cos\theta^{\prime}-\frac{2\alpha}{\pi}\sin\theta^{\prime}},
    \label{appeq:GML1}
\end{equation}
and when $\cos\theta^{\prime}-\frac{2\alpha}{\pi}\sin\theta^{\prime}<0$ we can have an upper bound
\begin{equation}
    \mathcal{F} \leq \frac{1-\frac{2t^{\alpha}}{\pi^{\alpha}}\langle(\hat{H}-E_{\min})^{\alpha}\rangle}{\cos\theta^{\prime}-\frac{2\alpha}{\pi}\sin\theta^{\prime}}.
\label{appeq:GML2}
\end{equation}
Similarly, we can also extract the dual generalized ML bound as below
\begin{equation}
    \mathcal{F} \geq \frac{1-\frac{2t^{\alpha}}{\pi^{\alpha}}\langle(E_{\max} - \hat{H})^{\alpha}\rangle}{\cos\theta^{\prime\prime}-\frac{2\alpha}{\pi}\sin\theta^{\prime\prime}},
\end{equation}
when $\cos\theta^{\prime\prime}-\frac{2\alpha}{\pi}\sin\theta^{\prime\prime}>0$ and 
\begin{equation}
    \mathcal{F} \leq \frac{1-\frac{2t^{\alpha}}{\pi^{\alpha}}\langle(E_{\max} - \hat{H})^{\alpha}\rangle}{\cos\theta^{\prime\prime}-\frac{2\alpha}{\pi}\sin\theta^{\prime\prime}},
    \label{appeq:GML4}
\end{equation}
when $\cos\theta^{\prime\prime}-\frac{2\alpha}{\pi}\sin\theta^{\prime\prime}<0$. Here we define $|\psi^{\prime\prime}(t)\rangle=e^{-{\rm i}(E_{\max} - \hat{H})t}|\psi(0)\rangle$ and $\langle\psi(0)|\psi^{\prime\prime}(t)\rangle = {\cal F} e^{-{\rm i}(E_{\max} - \hat{H})t} = {\cal F} e^{{\rm i}\theta^{\prime\prime}}$. Thus $\theta^{\prime\prime} = -(\theta + E_{\max} t)$. Combining Eqs.~\eqref{appeq:GML1} and \eqref{appeq:GML4} gives the generalized ML bound in the main text.

\section{Device information}
\begin{figure}[h]
    \begin{center}
    \includegraphics[width=1\textwidth]{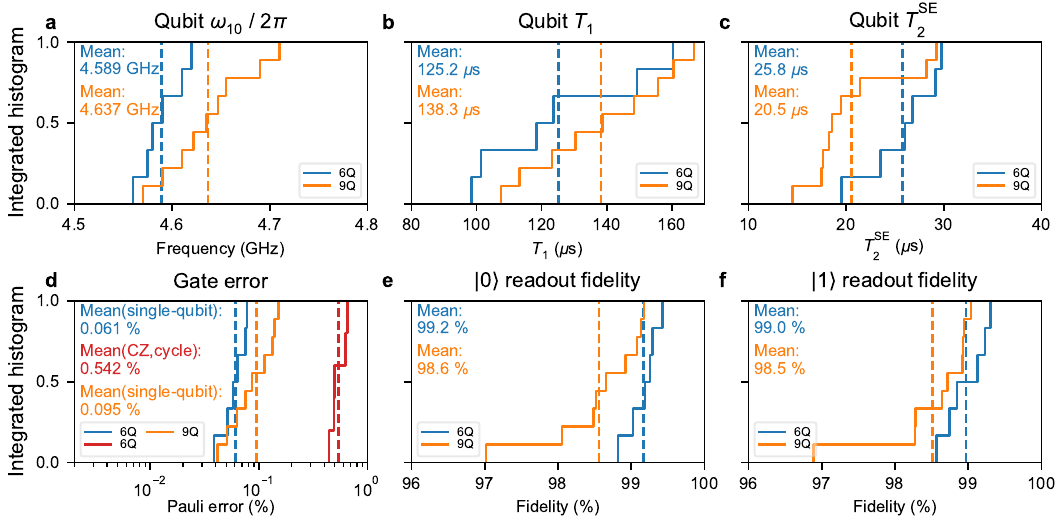}
    \end{center}
    \caption{{\bf Integrated histograms of typical qubit parameters for 6-qubit and 9-qubit systems.} Blue lines and orange lines represent the properties of 6-qubit system and 9-qubit system, respectively. Dashed lines denote mean values. {\bf a}, Statistics of qubit idle frequency $\omega_{10}/2\pi$. {\bf b}, Statistics of qubit energy relaxation time $T_1$. {\bf c}, Statistics of qubit spin echo dephasing time $T_2^{\rm{SE}}$. {\bf d}, Statistics of qubit gate error. The red line indicates the cycle Pauli errors of two-qubit CZ gates used in 6-qubit system. {\bf e}, Statistics of qubit state $|0\rangle$ readout fidelity. {\bf f}, Statistics of qubit state $|1\rangle$ readout fidelity.}
    \label{smfig:device}
\end{figure}
We perform the experiments on a superconducting quantum processor, the same as that in Ref.~\cite{ET_2022_NatPhys_FockSpaceDynamics, bao_2024_cat_dtc}, possessing 36 frequency-tunable transmon qubits and 60 tunable couplers. A $1\times 6$ one-dimensional (1d) chain and a $3\times 3$ two-dimensional (2d) square lattice 
are chosen to explore the quantum speed limits, respectively. The qubit and qutrit evolution are performed on the same qubit, also selected from the 1d chain and the 2d square lattice. Typical performance parameters of the qubits used in many-body systems are displayed in Fig.~\ref{smfig:device}. The idle frequencies $\omega_{10}/2\pi$ are shown in Fig.~\ref{smfig:device}{\bf a}. The energy relaxation time $T_1$ and spin echo dephasing time $T_2^{\rm{SE}}$ are shown in Fig.~\ref{smfig:device}{\bf b}, {\bf c}, with mean values of $T_1=125.2~\mu$s ($138.3~\mu$s) and $T_2^{\rm{SE}}= 25.8~\mu$s ($20.5~\mu$s) for 6-qubit (9-qubit) system. The single-qubit gate and two-qubit CZ gate cycle Pauli errors of the 6-qubit system are 0.061\% and 0.542\%, respectively, which are benchmarked with simultaneous cross-entropy benchmarking (XEB)~\cite{Boixo2018NP, Arute2019Nature}. Single-qubit Pauli error of the 9-qubit system is 0.095\%. Two-qubit gate is not used in the 9-qubit system. Qubit readout fidelities are shown in Fig.~\ref{smfig:device}{\bf e}, {\bf f}, whose average values reach 99.2\%, 99.0\%, 98.6\% and 98.5\% for measuring 6-qubit state $|0\rangle$, 6-qubit state $|1\rangle$, 9-qubit state $|0\rangle$ and 9-qubit state $|1\rangle$, respectively. 

For the qubit and qutrit evolution, we employ the same qubit but under different idle frequencies $\omega_{10}/2\pi=4.578$ GHz and 4.560 GHz, respectively. The nonlinearity $\eta/2\pi$ is calibrated as -212 MHz. The average energy relaxation time $T_1^{01}$ and $T_1^{12}$ over the two idle frequencies are 140.0~$\mu$s and 91.6~$\mu$s , respectively. Note that the superscripts `01' and `12' indicate the energy relaxation happens between state $|0\rangle$ and $|1\rangle$ or between state $|1\rangle$ and $|2\rangle$. Similarly, the average spin echo dephasing time $T_2^{\rm{SE}, 01}$~($T_2^{\rm{SE}, 12}$) is 25.8~$\mu$s~(21.6~$\mu$s). Single-qubit gate error $\epsilon_{\rm{SQ}}^{01}$~($\epsilon_{\rm{SQ}}^{12}$) is also benchmarked by XEB, with the average value of 0.038\%~(0.065\%). The readout fidelities used in qubit and qutrit evolution are different. For qubit case, the readout fidelities of state $|0\rangle$ and $|1\rangle$ are 99.3\% and 99.1\%. For qutrit case, the readout fidelities of state $|0\rangle$, $|1\rangle$ and $|2\rangle$ are 99.3\%, 95.8\% and 93.7\% .

\section{Calibration of microwave drive amplitude}

\begin{figure}[h]
    \begin{center}
    \includegraphics[width=1\textwidth]{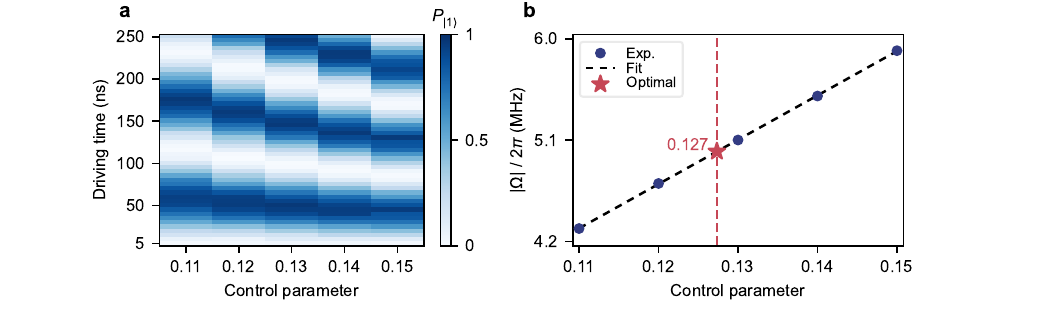}
    \end{center}
    \caption{{\bf Calibration of microwave drive amplitude.} 
    {\bf a}, Experimentally measured population dynamics up to 250 ns for five different control parameters of the microwave drive. 
    {\bf b}, The relationship between $\Omega$ and the experimental control parameter. By fitting the experimentally measured oscillations in {\bf a}, we can obtain the corresponding drive amplitudes (dark blue dots). We linearly fit the five experimental points and the estimated optimal drive parameter is 0.127 for $\Omega/2\pi=5$ MHz, which is indicated by the red star.}
    \label{smfig:cal_Omega}
\end{figure}

To map the microwave drive amplitude $\Omega$ in the Hamiltonian to experimental control parameter, we vary the control parameter to drive the qubit for a duration $t_d$ and then measure its population dynamics (Fig.~\ref{smfig:cal_Omega}{\bf a}) to fit the drive amplitude $\Omega$. The relationship between $\Omega$ and the control parameter is built by a linear fitting, from which we can choose the target drive amplitude accordingly (Fig.~\ref{smfig:cal_Omega}{\bf b}).

\section{Manipulation of three-level quantum system}
\begin{figure}[h]
    \begin{center}
    \includegraphics[width=1\textwidth]{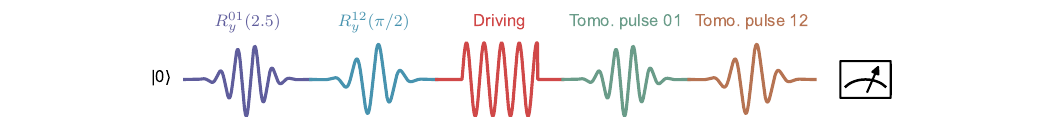}
    \end{center}
    \caption{{\bf Qutrit initial state preparation, evolution and tomography.} To prepare the state $\frac{1}{\sqrt{10}}[|0\rangle+\frac{3}{\sqrt{2}}(|1\rangle+|2\rangle)]$, first we apply the gate $R_y^{01}(2.5)$ to prepare the quantum state to $(1/\sqrt{10})(|0\rangle+3|1\rangle)$. Then the gate $R_y^{12}(\pi/2)$ is applied to rotate qutrit state around y-axis by $\pi/2$ in the Hilbert subspace of $|1\rangle$ and $|2\rangle$. After the initial state is prepared, the required driving pulse is applied. Note that 10 ns driving paddings are used before and after driving pulse. Finally, we apply qutrit tomography pulse mentioned in the text and measure the populations of state $|0\rangle$, $|1\rangle$ and $|2\rangle$ to construct the density matrix of qutrit.}
    \label{smfig:qutrit_circuit}
\end{figure}
To prepare the superposition state of a qutrit,  $|\psi(0)\rangle=\frac{1}{\sqrt{10}}[|0\rangle+\frac{3}{\sqrt{2}}(|1\rangle+|2\rangle)]$, we apply the quantum gates $R_y^{12}\left(\pi/2\right)R_y^{01}\left(2.5\right)$. In order to obtain the overlap $|\langle \psi(0)|\psi(t)\rangle|$, we need to implement quantum tomography for qutrit~\cite{Bianchetti_2010_PRL} (see Fig.~\ref{smfig:qutrit_circuit} for the gate sequence). 
Nine different operations are applied before the measurements: $I$, $R_x^{01}(\pi/2)$, $R_y^{01}(\pi/2)$, $R_x^{01}(\pi)$, $R_x^{12}(\pi/2)$, $R_y^{12}(\pi/2)$, $R_x^{12}(\pi/2)R_x^{01}(\pi)$, $R_y^{12}(\pi/2)R_x^{01}(\pi)$, $R_x^{12}(\pi)R_x^{01}(\pi)$. The superscript `01'~(`12') denotes the microwave frequency of the gate is $\omega_{10}$~($\omega_{21}$). The density matrix can be reconstructed by solving an over-determined equation using the measurement results.

\begin{figure}[h]
    \begin{center}
    \includegraphics[width=1\textwidth]{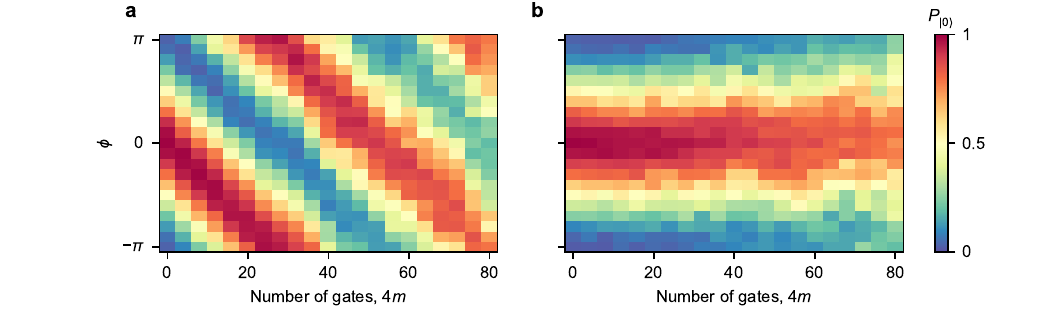}
    \end{center}
    \caption{{\bf Calibration of AC Stark shift phase $\bm{\phi_{\rm{ACS}}^{12\rightarrow 01}}$.} {\bf a}, Population without phase correction. The optimal phases for compensating $\phi_{\rm{ACS}}^{12\rightarrow 01}$ after $4m\cdot$$R_x^{12}(\pi)$ gates grow linearly with $m$ and can be obtained by fitting the cosine oscillation for each $m$. A linear fit between the optimal phase and number of gates $m$ gives the phase value for compensating $\phi_{\rm{ACS}}^{12\rightarrow 01}$ for each $R_x^{12}(\pi)$ gate. {\bf b}, Population with the phase correction obtained in {\bf a}.}
    \label{smfig:ac_phase}
\end{figure}

A crucial point for accurately manipulating a qutrit is to cancel the effect of AC Stark shift caused by the driving pulse~\cite{Morvan_2021_acphase}. It will causes an unwanted phase $\phi_{\rm{ACS}}^{12\rightarrow 01}$~($\phi_{\rm{ACS}}^{01\rightarrow 12}$) for 01-~(12-) level when a 12-~(01-) gate is applied. Thus, we apply a virtual $Z^{01}$~($Z^{12}$) gate~\cite{McKay_2017_vz} after each 12-~(01-) gate to compensate $\phi_{\rm{ACS}}^{12\rightarrow 01}$~($\phi_{\rm{ACS}}^{01\rightarrow 12}$). For example, the  $\phi_{\rm{ACS}}^{12\rightarrow 01}$ can be calibrated as the following. First, we apply a $R_y^{01}(\pi/2)$ gate to qutrit to prepare it to  $(1/\sqrt{2})(|0\rangle+|1\rangle)$. Then $4m\cdot$$R_x^{12}(\pi)$ gates, where $m\in\mathbb{N}$, are applied to accumulate $\phi_{\rm{ACS}}^{12\rightarrow 01}$ for $4m$ times. Finally a virtual Z gate $R_z^{01}(\phi)$ followed by a recovery gate $R_y^{01}(-\pi/2)$ is employed. By sweeping $\phi$ and measuring the population of $|0\rangle$, we can find the optimal phase we should use. Similarly, we can exchange all 01- and 12- gates and measure the population of $|1\rangle$ to calibrate the optimal phase for compensating $\phi_{\rm{ACS}}^{01\rightarrow 12}$. Note that a $R_y^{01}(\pi)$ gate is needed before the calibration starts to pump the qutrit to $|1\rangle$. The calibration results before and after correcting $\phi_{\rm{ACS}}^{12\rightarrow 01}$ are depicted in Fig.~\ref{smfig:ac_phase}. 

\section{Longer time dynamics of the three-level quantum system}
\begin{figure}[h]
    \begin{center}
    \includegraphics[width=1\textwidth]{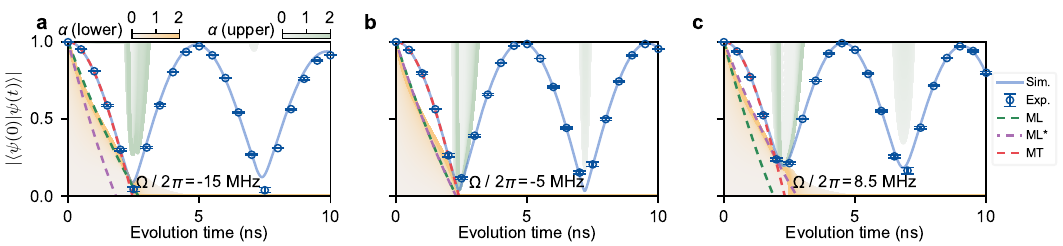}
    \end{center}
    \caption{{\bf Quantum speed limit for qutrit.} {\bf a}, {\bf b}, {\bf c},  Evolutions for $\Omega/2\pi=-15$ MHz, -5 MHz and 8.5 MHz. The blue circles (lines) are the experimental (theoretical) results of $|\langle\psi(0)|\psi(t)\rangle|$; dashed lines indicate the ML, $\rm ML^{\star}$ and MT bounds. Error bars are computed by repeating measurement for 5 times. Here, we display a longer time scale up to 10 ns to show the generalized bounds also constrain the second oscillation period.}
    \label{smfig:qutrit_subfig}
\end{figure}
Figure~\ref{smfig:qutrit_subfig} shows the overlap dynamics of the qutrit over a much longer time scale than those in Fig.~2 of the main text. Here, the key point is to show the periodic nature of the overlap dynamics, a direct consequence of the closed orbits in the generalized Bloch sphere, and that the generalized bounds constraints also apply to the second oscillation period, albeit with a less stringent bound.

\section{Control and measurement of coupling strengths}

Observation of quantum speed limits requires to control and measure the coupling strengths between qubits precisely. Our chip is a rectangular lattice, on which we only consider the actively controllable nearest-neighbor coupling and uncontrollable cross-coupling.

We can individually control nearest-neighbor coupling strengths for different qubit pairs by utilizing tunable couplers~\cite{Yan_2018_coupler}. The effective coupling strength between a qubit pair is
\begin{equation}
    \tilde{g}\approx \frac{g_1 g_2}{2}(\frac{1}{\Delta_1}+\frac{1}{\Delta_2})+g_{12},  
\end{equation}
where $g_i$ is the coupling strength between coupler and the qubit $i$ and $g_{12}$ the capacitive coupling strength of the two qubits. $\Delta_i=\omega_i-\omega_c$ is the detuning of the qubit $i$ from the coupler. By adjusting $\Delta_1$ and $\Delta_2$, the effective coupling strength $\tilde{g}/2\pi$ can be dynamically tuned up to about -20 MHz.

As for measuring coupling strengths, we follow the procedure described in Ref.~\cite{Xiang_2024_QST}. For nearest-neighbor couplings, we keep each qubit pair at the interaction frequency $\omega_{\rm{I}}$, making all other qubits detuned $\Delta/2\pi=\pm 120 $ MHz (or $\pm 70$ MHz for one qubit limited by its maximum frequency) away from $\omega_{\rm{I}}$ at the same time. For cross-couplings, we do the same for a set of detunings $\Delta/2\pi\in\{-150~\rm{MHz}, -120~\rm{MHz}, -90~\rm{MHz}, -60~\rm{MHz}, 50~\rm{MHz}, 70~\rm{MHz}\}$ to get a set of effective coupling strengths, which can be used to fit the actual cross-coupling strengths. 

\section{Preparation of 6-qubit superposition state}

\begin{figure}[h]
    \begin{center}
    \includegraphics[width=1\textwidth]{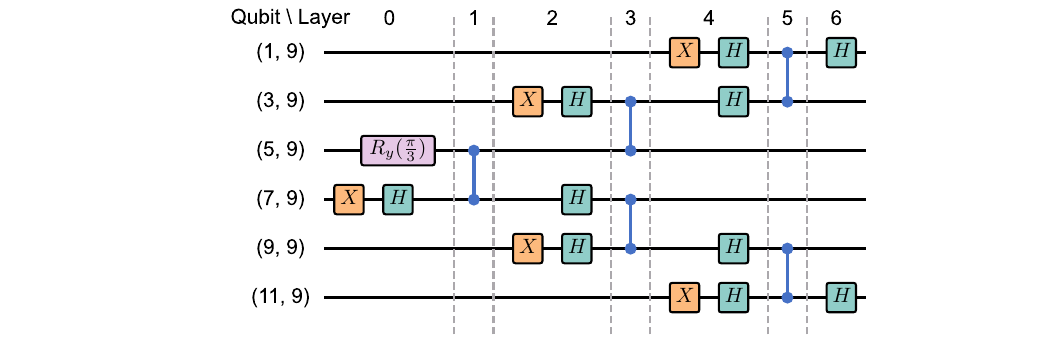}
    \end{center}
    \caption{{\bf Quantum circuit for preparing superposition state.} The circuit consists of 7 layers, including 3 layers of two-qubit CZ gates and 4 layers of single-qubit gates. Single-qubit gates or two-qubit CZ gates are applied simultaneously to the corresponding qubits in each layer. Global entanglement is constructed gradually by applying CZ gates layer by layer.}
    \label{smfig:cat_circuit}
\end{figure}

The quantum circuit for generating 6-qubit superposition state $\left(\sqrt{3}/2\right)|010101\rangle+\left(1/2\right)|101010\rangle$ is similar to Ref.~\cite{bao_2024_cat_dtc}, as depicted in Fig.~\ref{smfig:cat_circuit}. The basic idea is to initialize the first qubit to $\left(\sqrt{3}/2\right)|0\rangle+\left(1/2\right)|1\rangle$ and spread the entanglement by applying CNOT gates, which are decomposed into CZ gates and Hadamard gates. Note that any coefficient of the two components is accessible as long as we substitute the corresponding single-qubit gate $R_y(\pi/3)$.

\section{Measurement of the overlap with superposition initial state}

The overlap between a state $|\psi(t)\rangle$ at time $t$ and the initial superposition state $|\psi_{\rm{cat}}\rangle=\left(\sqrt{3}/2\right)|0101\cdots\rangle+\left(1/2\right)|1010\cdots\rangle$ is defined as $|\langle \psi(t)|\psi_{\rm{cat}}\rangle|$, which requires to measure the diagonal elements $\{\rho_{1010\cdots}$, $\rho_{0101\cdots}\}$ and off-diagonal elements $\{\rho_{1010\cdots, 0101\cdots}$, $\rho_{0101\cdots, 1010\cdots}\}$ of the density matrix. The former can be measured directly while obtaining the latter needs to measure parity oscillations. Parity oscillations~\cite{sackett_2000_parity, song_2019_ghz, bao_2024_cat_dtc} are commonly used to probe the off-diagonal elements. Parity is given by
\begin{subequations}
\begin{align}
    \langle\mathcal{P}(\gamma)\rangle &= \left\langle\bigotimes_{j=1}^N \left(\cos((-1)^j\gamma)\sigma^y_j+\sin((-1)^j\gamma)\sigma^x_j\right)\right\rangle\\
    &=(-i)^N\left\langle\bigotimes_{j=1}^N \left(\begin{matrix}
0&e^{i(-1)^j\gamma}\\-e^{-i(-1)^j\gamma}&0\end{matrix}\right)\right\rangle,
\end{align}
\end{subequations}
where $N$ is the number of qubits, and $\gamma$ is a variable parameter. For any $N$-qubit superposition state 
\begin{equation}
   |\psi\rangle = \cos\frac{\theta}{2}|0101\cdots\rangle+e^{i\phi}\sin\frac{\theta}{2}|1010\cdots\rangle,
   \label{smeq:cat_state}
\end{equation}
where $\theta\in [0,\pi]$ and $\phi\in [0,2\pi]$, parity has the form
\begin{align}
   \langle\mathcal{P}(\gamma)\rangle =
   \begin{cases}
      \sin\theta \sin(N\gamma+\phi), & {\rm if~}N \rm{~is~odd}\ ,\\    
      \sin\theta \cos(N\gamma+\phi), & {\rm if~} N {\rm~is~even}.
    \end{cases}
    \label{smeq:parity}
\end{align}
Note that $\rho_{1010\cdots, 0101\cdots}=\sin\theta e^{i\phi}/2$ and $\rho_{0101\cdots, 1010\cdots}=\sin\theta e^{-i\phi}/2$. $\theta$ and $\phi$ can be determined by fitting parity oscillations as a function of $\gamma$. Then, the off-diagonal elements can be known.

\begin{figure}[t!]
    \begin{center}
    \includegraphics[width=1\textwidth]{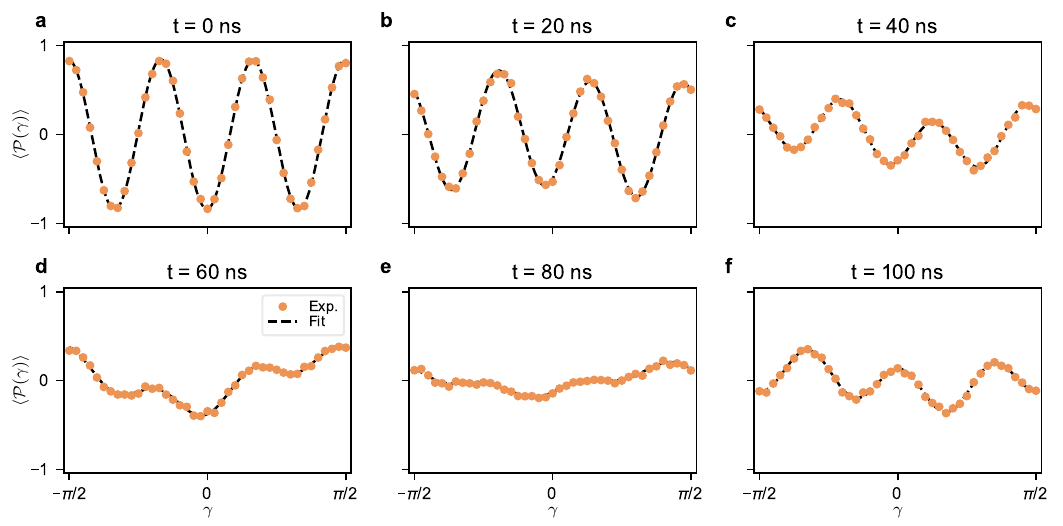}
    \end{center}
    \caption{{\bf Snapshots of parity oscillations for \bm{$W/2\pi=1.8$} MHz at different evolution time \bm{$t$}.} {\bf a}-{\bf f} show the snapshots of parity oscillations chosen from 0 ns to 100 ns evenly. Orange circles indicate the experimentally measured data. Black dashed lines are fitted using Eq.~\eqref{smeq:tri_series} for $N=6$.}
    \label{smfig:parity}
\end{figure}

Since we need to measure the overlap between $|\psi_{\rm{cat}}\rangle$ and other quantum states (i.e., $|\psi(t)\rangle$), not only the prepared initial state, it is insufficient to fit the parity curve using Eq.~\eqref{smeq:parity}. The fitting formula can be generalized to
 \begin{equation}
     \langle\mathcal{P}(\gamma)\rangle=
     \begin{cases}
      \displaystyle\sum_{n\in\{1, 3, \cdots,N\}} c_n\sin(n\gamma+\xi_n), & {\rm if~}N {\rm~is~odd}\ , \\
      \displaystyle\sum_{n\in\{0, 2, \cdots,N\}} c_n\cos(n\gamma+\xi_n), & {\rm if~}N {\rm ~is~even},
    \end{cases}
    \label{smeq:tri_series}
\end{equation}
where $c_n$ and $\xi_n$ are the amplitude and the phase, respectively. Then, the off-diagonal elements can be extracted similarly:
\begin{subequations}
\begin{align}
    \rho_{1010\cdots, 0101\cdots}&=\frac{1}{2}c_N e^{i\xi_N}\ ,\\
    \rho_{0101\cdots, 1010\cdots}&=\frac{1}{2}c_N e^{-i\xi_N}.
\end{align}
\end{subequations}

By fitting the experimentally measured parity oscillations, as illustrated in Fig.~\ref{smfig:parity}, we can obtain the overlap $|\langle \psi(t)|\psi_{\rm{cat}}\rangle|$. This method avoids the time-consuming 6-qubit quantum state tomography.

\section{Effects of imperfect pulse control for qubit and qutrit}
\begin{figure}[h]
    \begin{center}
    \includegraphics[width=0.8\textwidth]{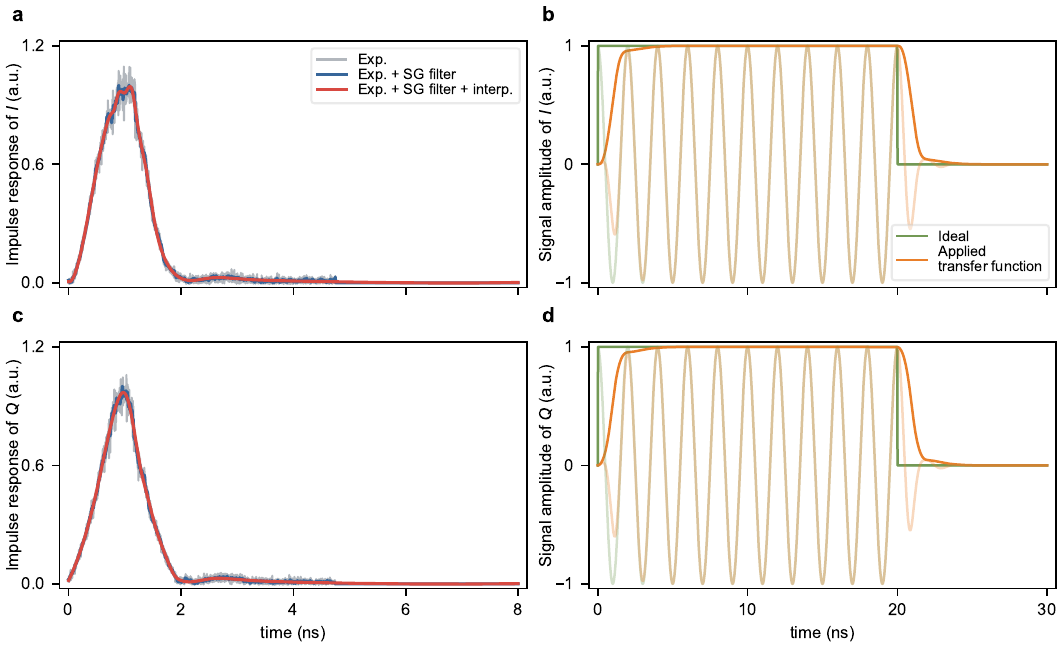}
    \end{center}
    \caption{{\bf Impulse response and the rectangular pulse with rising and falling edge for qutrit simulation.}
    {\bf a} and {\bf c} show the impulse response of two channels $I$ and $Q$ of AWG, respectively. We use a sampling oscilloscope to probe the response of an 0.5 ns impulse (single point of AWG). Here, the frequency of carrier wave is 4.8 GHz. Hilbert transform is implemented to extract the envelope of the output signal~(gray lines). To obtain a smooth impulse response, Savitzky–Golay filter~\cite{savitzky_1964_sg_filter} and spline interpolation are applied to sampled data successively, corresponding to blue and red lines, respectively.
    {\bf b} and {\bf d} display the ideal driving pulse~(green lines) and the distorted driving pulse~(orange lines). We multiply ideal signal by smoothed impulse response in frequency domain and then implement inverse fast Fourier transform~(IFFT) to compute the distorted time-domain signal. Then we can use the pulse that is closer to the realistic situation to simulate the time evolution of the driven single-qubit and single-qutrit Hamiltonians.}
    \label{smfig:pulse_shape}
\end{figure}

For qubit or qutrit Hamiltonian described by Eq.~(1) or (3) of the main text, a driving pulse with a rectangular envelope is required. However, it is infeasible due to the finite rising and falling edge of the arbitrary waveform generator~(AWG). To investigate the influence of imperfect pulse shape, we use a sampling oscilloscope to probe the output signals of AWG. Two channels $I$ and $Q$ are needed to output the real part and the imaginary part of a complex pulse. In practice, measuring the impulse response (see Fig.~\ref{smfig:pulse_shape} for experimental data) suffices to obtain the distorted output of any ideal pulse input.

\begin{figure}[t!]
    \begin{center}
    \includegraphics[width=0.8\textwidth]{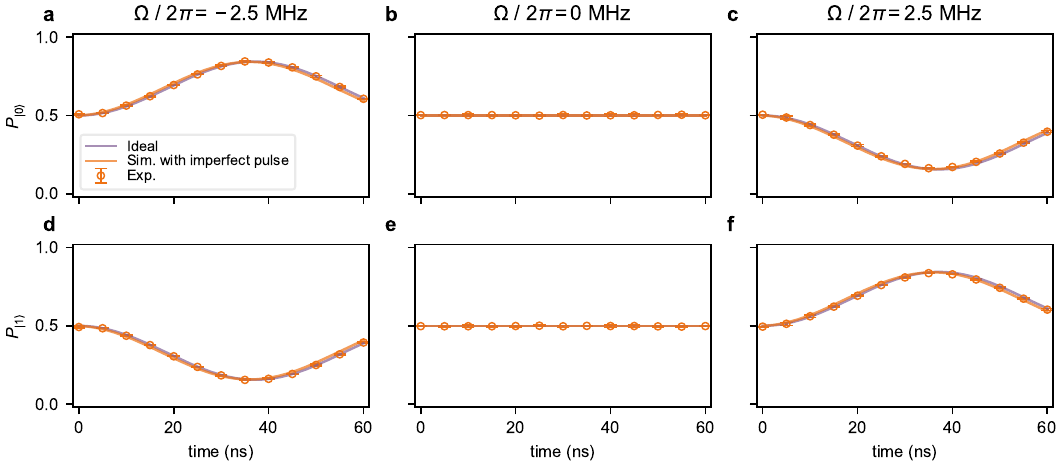}
    \end{center}
    \caption{{\bf Population dynamics of state $|0\rangle$ and $|1\rangle$ for different drive amplitudes.}
    Purple lines are ideal simulation results, i.e., the envelope of pulse is a perfect rectangular pulse and the maximum energy level is 2. Orange lines denote the simulation results by applying the transfer function in Fig.~\ref{smfig:pulse_shape}{\bf a} and {\bf c} to ideal pulse and considering 4 lowest energy levels instead of 2. Orange circles represent experimental results. Error bars are computed from 5 repetitions of  measurements.
    {\bf a}, {\bf d}, Population dynamics for  $\Omega/2\pi=-2.5$ MHz.
    {\bf b}, {\bf e}, Population dynamics for  $\Omega/2\pi=0$ MHz.
    {\bf c}, {\bf f}, Population dynamics for  $\Omega/2\pi=2.5$ MHz. 
    }
    \label{smfig:qubit_probs}
\end{figure}

\begin{figure}[t!]
    \begin{center}
    \includegraphics[width=0.8\textwidth]{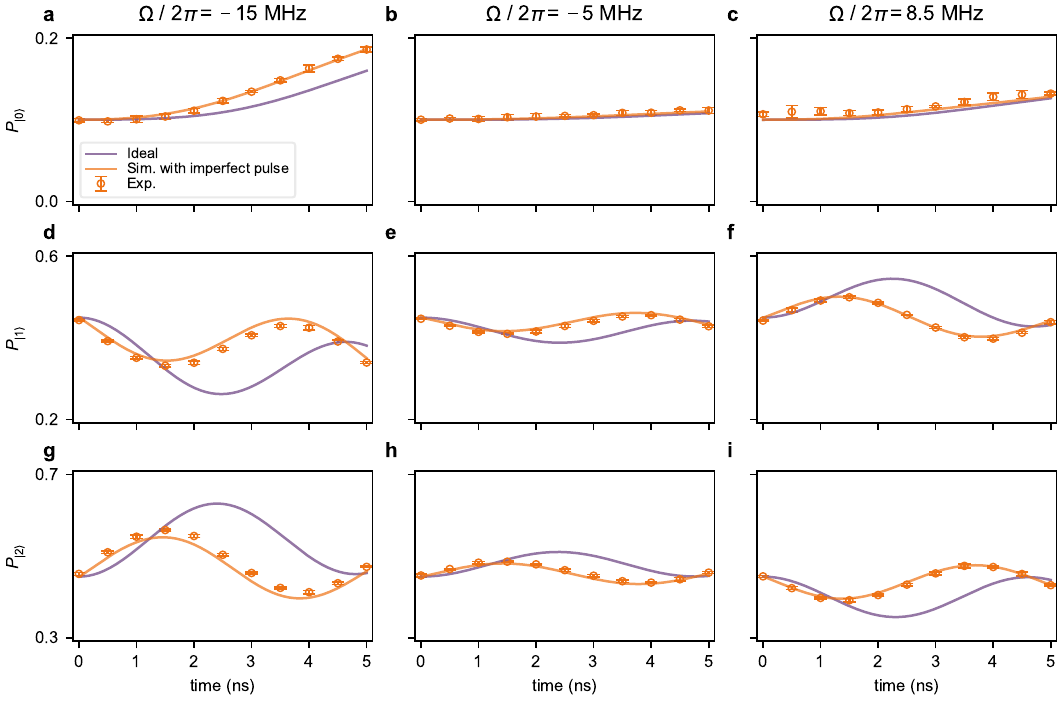}
    \end{center}
    \caption{{\bf Population dynamics of state \bm{$|0\rangle$}, \bm{$|1\rangle$} and \bm{$|2\rangle$}  for different drive amplitudes.}
    Purple lines are ideal simulation results, i.e., the envelope of pulse is a perfect rectangular pulse and the maximum energy level is 3. Orange lines denote the simulation results by applying the transfer function in Fig.~\ref{smfig:pulse_shape}{\bf a} and {\bf c} to ideal pulse and considering 4 lowest energy levels instead of 3. Orange circles represent experimental results. Error bars are computed from 5 repetitions of measurements.
    {\bf a},{\bf d},{\bf g}, Population dynamics for $\Omega/2\pi=-15$ MHz.
    {\bf b},{\bf e},{\bf h}, Population dynamics for $\Omega/2\pi=-5$ MHz.
    {\bf c},{\bf f},{\bf i}, Population dynamics for $\Omega/2\pi=8.5$ MHz.
    }
    \label{smfig:qutrit_probs}
\end{figure}

With the measured impulse response, we can calculate the distorted pulse and then use it to perform the numerical simulations to explain the difference between experimental and ideal simulation results. Figs.~\ref{smfig:qubit_probs} and \ref{smfig:qutrit_probs} show the populations during the evolution of qubit and qutrit, respectively. For the dynamics of qutrit, there is a notable difference between the ideal numerical and experimental results. It can be explained if we use the distorted pulse to do the simulation. Note that if we want to simulate the population at time $t$, we need to append a driving-padding $\Delta t$ behind $t$ to make the transformed pulse fully takes the effect. Here, we choose $\Delta t=10$ ns.

\section{Simulation of larger 2d square lattice systems}
\begin{figure}[h]
    \begin{center}
    \includegraphics[width=0.8\textwidth]{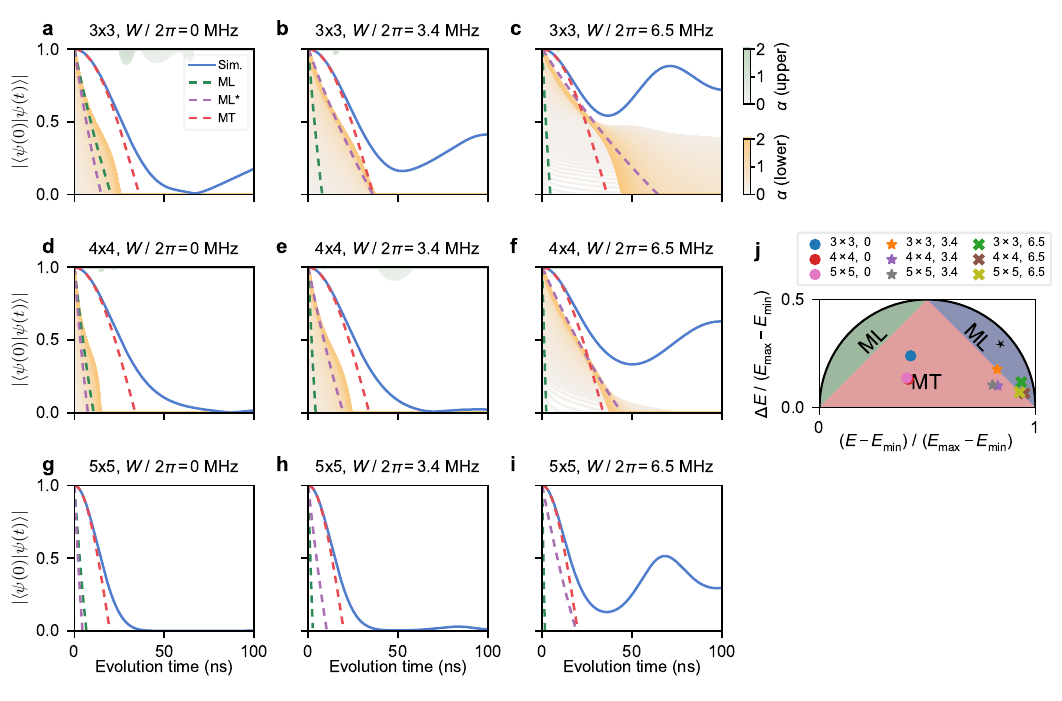}
    \end{center}
    \caption{{\bf Numerical results for different lattice size systems.} {\bf a}-{\bf i}, Dynamics for the overlap and corresponding bounds for $3\times 3$, $4\times 4$, and $5\times 5$ systems. A checkerboard potential is included in all cases, with initial excitations in the $+W$ sites. This leads to 5 excitations in the $3\times 3$ case (Hilbert space dimension ${\cal D} = 126$), 8 excitations in the $4\times 4$ (${\cal D} = 12,870$) and 13 excitations in the $5\times 5$ (${\cal D} = 5,200,300$). Color plots in {\bf a}-{\bf f} show the generalized bounds' results, and various colors represent the magnitude of the exponent $\alpha$. Other parameters are the same as in Fig.~4 in the main text. {\bf j}, Phase diagram with markers annotating the loci of the remaining panels and their various parameters.
    }
    \label{smfig:larger_sim}
\end{figure}

In the main text, we argue that in the case of many-body quantum Hamiltonians, the unified orthogonality time $t_\perp\geq t_{\rm U}=\max\left(t_{\rm MT}, t_{\rm ML}, t_{\rm ML^{\star}}\right)$ is guaranteed to be reduced once one takes the thermodynamic limit while preserving the ratio of the number of excitations to the number of qubits. Behind this is that the physical quantities that determine this minimal time for orthogonalization, $\Delta E$ and $E$, are extensive in the system size; 
consequently, the bounds ought to become less restrictive. For instance, in Fig.~\ref{smfig:larger_sim}, deep in the MT regime with $W/2\pi=0$ MHz, $t_{\rm MT}$ substantially decreases while growing the two-dimensional lattice from $9$ qubits to 25. However, since the Hilbert space dimension grows roughly $41000$-fold, one expects that the decay rate of the overlap will also be much faster in this richer vectorial space. In practice, this leads to bounds on the dynamics before $t_{\rm MT}$, which are as tight as before, happening just in shorter time scales.

Second, finite-size effects can result in qualitative different effects: While $W/2\pi = 3.4$ MHz leads to time bounds $t_{\rm MT} = t_{{\rm ML}^*}$ on a $3\times 3$ qubit network (Fig.~\ref{smfig:larger_sim}{\bf b}), $t_{\rm MT}$ is the relevant orthogonalization time if growing the system size (Fig.~\ref{smfig:larger_sim}{\bf e} and {\bf h}). The opposite is also observed, i.e., reducing the system size can lead to dynamics with bounds  $t_{\rm MT} = t_{{\rm ML}^*}$ to $t_{\rm MT} < t_{{\rm ML}^*}$ for fixed $W/2\pi = 6.5$ MHz (Fig.~\ref{smfig:larger_sim}{\bf c}, {\bf f} and {\bf i}). Lastly, we notice that the generalized bounds are similarly affected by extensive behaviors. Due to the difficulty of extracting the generalized bounds for the 25-qubit system, only 9- and 16-qubit results are exhibited here. One can see that larger sizes make the generalized bounds less tight for a fixed $W$. However, the generalized bounds can still surpass the other standard three after the crossover points when going into the ML$^{\star}$ regime (Fig.~\ref{smfig:larger_sim}{\bf c} and {\bf f}), emphasizing their generic validity for larger system sizes. On a technical aspect, we notice that the computation of the generalized bounds [Eqs.~(5) and (6) in the main text] entail the computation of the (time-independent) fractional powers of $\langle \hat H^\alpha\rangle$ ($\alpha >0$). For that, we utilize an algorithm based on Pad\'e approximants after a Schur decomposition of the original matrix~\cite{Higham2011}. Usual methods based on either eigendecomposition or using alternative expressions such as $\langle \hat H^\alpha\rangle = \langle \exp[\alpha \log (\hat H)]\rangle$ have shown to be less accurate and stable, in particular for large Hilbert space sizes and small values of $\alpha$.

\section{Integrable case: 1d chain}
Integrable systems like the one given by Eq.~(7) in the main text are known to evade fully chaotic behavior with the emergence of periodic revivals in physical quantities -- their thermalization is described by a generalized mechanism in which not only energy and number of particles are conserved but also a full set of one-body conserved quantities established by the occupancies of single-particle eigenstates of the corresponding fermionic Hamiltonian it can be mapped onto~\cite{Cassidy2011}. For that, one can observe that spin-1/2 operators are mapped into hardcore bosons: $\hat \sigma_{m}^+ \leftrightarrow \hat a_m^\dagger $ and $\hat \sigma_{m}^- \leftrightarrow \hat a^{\phantom{\dagger}}_m $~\cite{Matsubara1956}; the latter satisfying the usual bosonic commutation relations at different sites:
\begin{equation}
    [\hat a^{\phantom{\dagger}}_m, \hat a^\dagger_n] = [\hat a^{\phantom{\dagger}}_m, \hat a^{\phantom{\dagger}}_n] = [\hat a^\dagger_m, \hat a^\dagger_n]=0\ {\rm for\ }m\neq n
\end{equation}
while obeying
\begin{equation}
    \{\hat a_m^{\phantom{\dagger}}, \hat a_m^\dagger\} = 1,\ \hat a_m^{\dagger\ 2} = \hat a_m^{ 2} = 0\ ,
\end{equation}
relations typical of fermions on the \textit{same} site. Now, making use of the Jordan-Wigner transformation,
\begin{equation}
    \hat a_m^{\phantom{\dagger}} = \prod_{\beta=1}^{m-1} e^{{\rm i} \pi \hat c_{\beta}^\dagger \hat c_{\beta}^{\phantom{\dagger}}}\hat c_{m}^{\phantom{\dagger}}, \quad  \hat a_m^\dagger = \hat c_{m}^\dagger\prod_{\beta=1}^{m-1} e^{-{\rm i} \pi \hat c_{\beta}^\dagger \hat c_{\beta}^{\phantom{\dagger}}}\ ,
    \label{smeq:jw}
\end{equation}
the original Hamiltonian is recast in terms of non-interacting spinless fermions:
\begin{equation}
    \hat{H}_{\rm 1d}/\hbar=J_1\sum_{\avg{mn}}(\hat{c}_m^\dagger \hat c_n^{\phantom{\dagger}}+\hat c_n^{\dagger}\hat c _m^{\phantom{\dagger}})+\sum_m W_m\hat c_m^\dagger \hat c_m^{\phantom{\dagger}}\ .
\label{smeq:H_fermion} 
\end{equation}
With that, the dynamics of large system sizes can be easily computed based on the evolution of Slater determinants~\cite{Rigol2005}, together with the computation of relevant observables, such as the single-particle density matrix $\rho_{mn} \equiv \langle \hat a_m^\dagger \hat a_n^{\phantom{\dagger}}\rangle$. Its Fourier transform, the momentum distribution, $n_k = (1/L)\sum_{m,n} e^{{\rm i}k(m-n)} \rho_{mn}$, can be used to quantify (quasi-) condensation.

\begin{figure}[h]
    \begin{center}
    \includegraphics[width=0.7\textwidth]{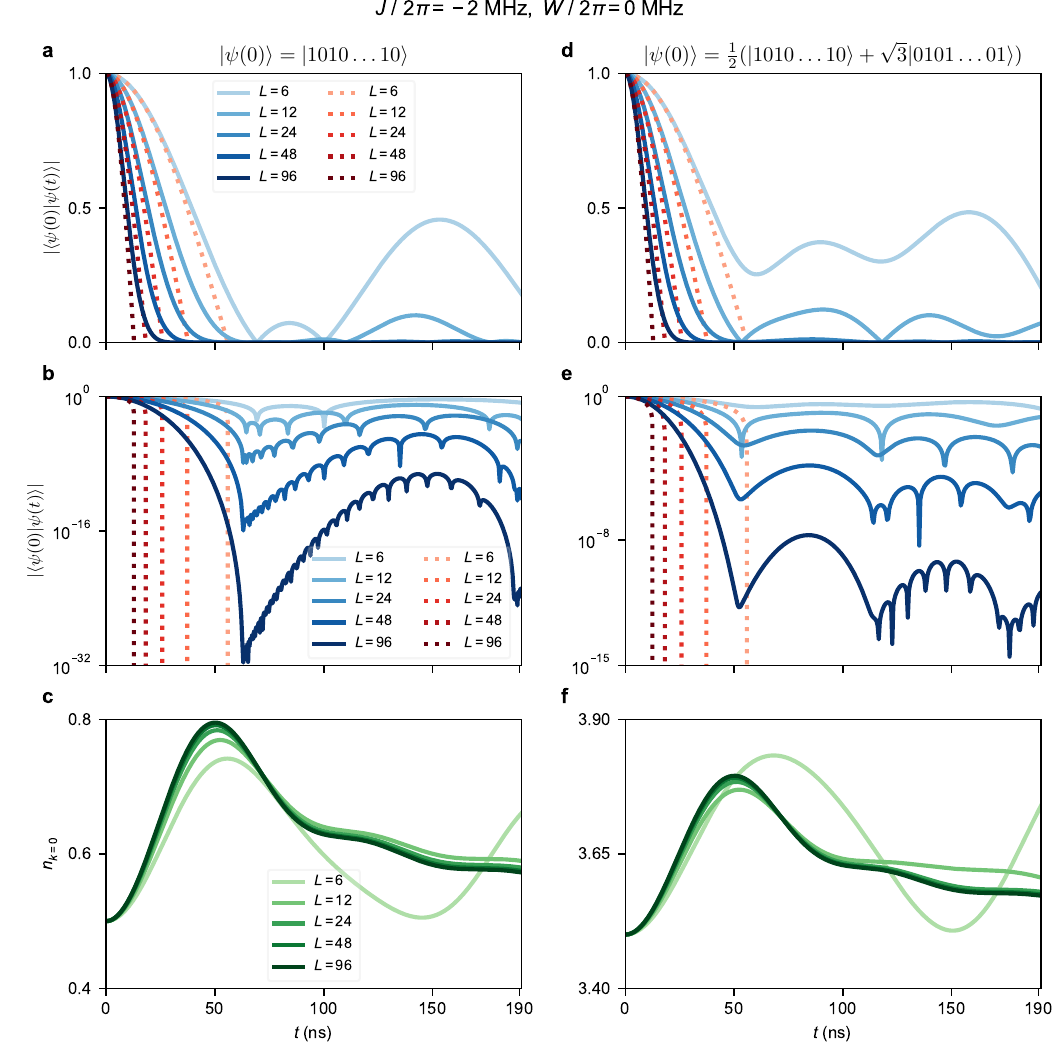}
    \end{center}
    \caption{{\bf Numerical results for different sizes of chain systems.} {\bf a}, The dynamics of overlap starting from a density-wave, product state $|\psi(0)\rangle = |1010\ldots 10\rangle$ for the one-dimensional Hamiltonian [Eq.~(7) in the main text] with $W/2\pi=0$ MHz; the system sizes are $L=6, 12, 24, 48$ and 96; the dotted lines are the corresponding MT bounds. {\bf b}, The same but in logarithmic scale, highlighting the orthogonalization times. {\bf c}, Dynamics of the zero-momentum occupancy for the different system sizes; its first peak is not aligned with the peak of the zero-momentum occupancy, which signals the initial build-up of one-body correlations. The bounds on the overlap can be easily computed for this Hamiltonian on our product state by noticing that the time-independent quantities $\langle \hat H_{\rm 1d} \rangle = 0$ and that $\langle \hat H_{\rm 1d}^2 \rangle =\langle \hat H_{\rm 1d} \hat H_{\rm 1d}\rangle$ can be computed by counting the number of possible hopping terms $\langle \hat H_{\rm 1d}^2 \rangle = J^2 [2(L/2-1)+1]=J^2(L-1)$ on a 1d lattice featuring $L$ qubits (sites). {\bf d}, {\bf e} and {\bf f} show the same for the superposition of product states used in the experiments: $|\psi(0)\rangle = \frac{1}{2}(|1010\ldots 10\rangle + \sqrt{3} |0101\ldots 01\rangle)$; similarly $\langle \hat H_{\rm 1d} \rangle = 0$ and $\langle \hat H_{\rm 1d}^2 \rangle = J^2(L-1)$ (for $L>4$) and the orthogonalization time does not align with the peak of $n_{k=0}$.}
    \label{smfig:larger_sim_1d}
\end{figure}

Our goals here are to (i) check the tightness of the QSL bounds in large system sizes and (ii) verify if physical quantities can capture the orthogonalization time $t_{\perp}$. Figure~\ref{smfig:larger_sim_1d} focuses on the regime with $W/2\pi =0$ MHz, where the MT bound is most relevant (see Fig.~3{\bf b} in the main text). Regarding the first goal above, increasing the system size leads to shorter bounds on the orthogonalization time $t_{\rm MT}$. Still, similar to the non-integrable case (Fig.~\ref{smfig:larger_sim}), there is a faster decay of the overlap such that the bounds are always constraining the dynamics for $t < t_{\rm MT}$. Regarding the second goal, we notice that the zero-momentum ($k=0$) occupancy, $n_{k=0}$, leads to the formation of peaks at short time scales and, subsequently, a slow decay towards the generalized Gibbs ensemble prediction for the equilibration. This initial peak is traced back from the build-up of one-body correlations of the initial uncorrelated product state~\cite{Rigol2006}. While the dips in the overlap signaling orthogonalization (see Fig.~\ref{smfig:larger_sim_1d}{\bf b} for the overlap in logarithmic scale) occurs at roughly $60$ ns, with a small size dependence, the peaks on $n_{k=0}$ are at slightly shorter time-scales. This exposes that zeros of overlap are purely many-body effects and cannot be explained by few body-correlators, even if these fundamentally characterize the dynamics.

\begin{figure}[h]
    \begin{center}
    \includegraphics[width=0.8\textwidth]{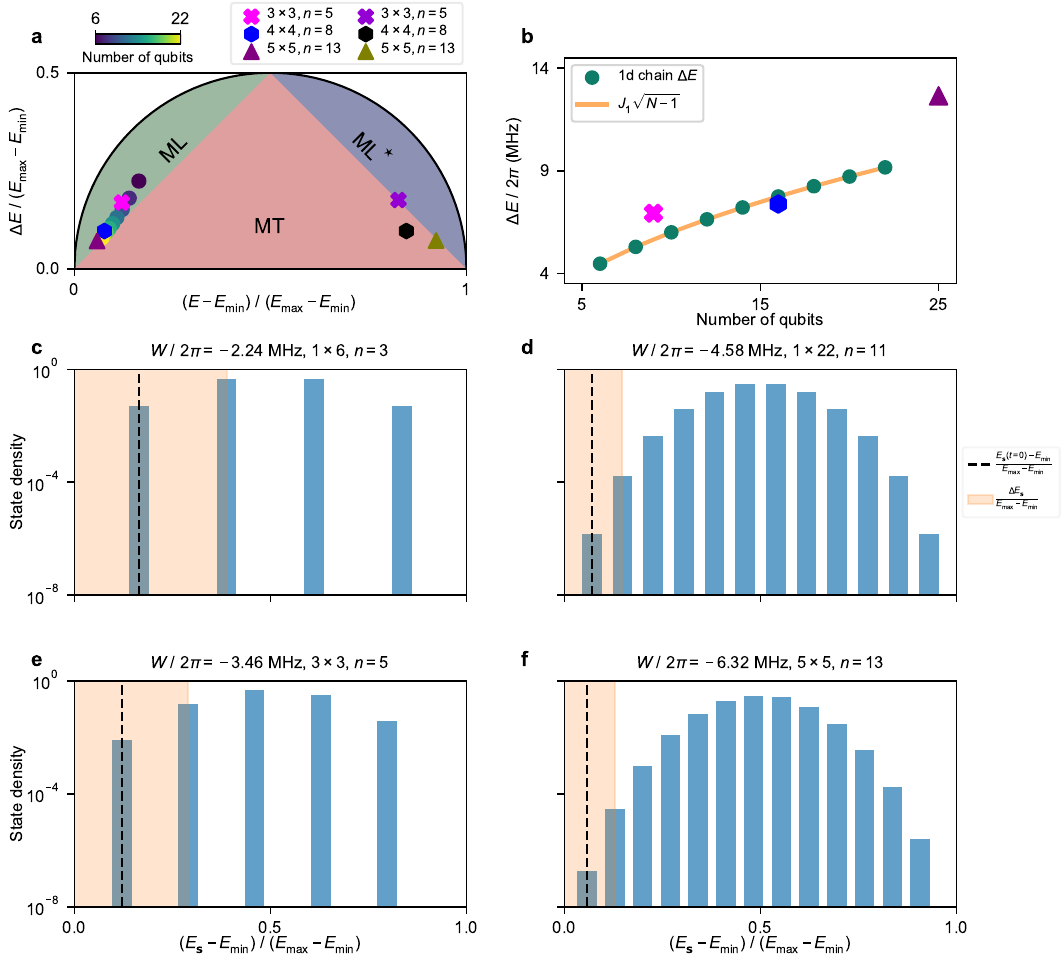}
    \end{center}
    \caption{{\bf Comparison of the positions in the phase diagram of 1d and 2d.} {\bf a}, The positions of different system sizes of 1d chain (filled circles) and 2d square lattice (other markers) by setting $\Delta E=2|W|, W<0$. $\Delta E$ is $W$-independent under the configuration of the Fock initial states $|1010\ldots\ 10\rangle$ and checkerboard on-site potentials $|W,-W,W,-W,\ldots,W,-W\rangle$ (see Section~\ref{appsec:comparison} for more details). The colorbar maps the number of qubits of the 1d chain ($N=6,8,\ldots,22$). The left and right cross markers depict the results of $3\times3$ square lattice of the same magnitude of $W$ but the opposite sign (the left is negative, and the right is positive). Hexagons and triangles are the corresponding results of $4\times4$ and $5\times5$. With increasing system sizes, the position gets close to the boundary of the MT and ML regimes. {\bf b}, Dependence of $\Delta E$ with the number of qubits. The green circles are the numerical results, and the solid line is the analytical representation. The other three markers have the same meaning as {\bf a}. {\bf c}-{\bf f}, Energy of Fock states and corresponding density of states (similar to Fig.~4{\bf g} - {\bf i} of the main text). The vertical dark orange line marks the mean energy $E$, and the shaded area represents the variance.}
    \label{smfig:compare_sizes}
\end{figure}
\section{Scaling of the positions in the phase diagram}
\label{appsec:comparison}

In Fig.~4 of the main text, we show the surprising result that when $\Delta E$ is close to the energy difference of adjacent Fock states, which exactly equals $2W$ for the special setup of the initial Fock state and on-site potentials we study, it marks the boundary between MT and ML$^{\star}$ bounds. A natural question is whether this is a generic conclusion. To answer this question, we first show that under the same setup of the initial density-wave Fock state, $\Delta E$ is $W$-independent. First, given the concrete forms of the 1d and 2d Hamiltonian [Eqs.~(7) and (8) in the main text], we can rewrite them as $\hat{H}=\hat{H}_{\rm hopping} + \hat{H}_{\rm local}$; a sum of off-diagonal and diagonal operators in the Fock basis. Computing $\Delta E^2=\avg{\hat{H}^2}-\avg{\hat{H}}^2$, leads to,
\begin{equation}
    \Delta E^2 = \avg{\hat{H}_{\rm hopping}^2}+\avg{\hat{H}_{\rm hopping}\hat{H}_{\rm local}}+\avg{\hat{H}_{\rm local}\hat{H}_{\rm hopping}}+\avg{\hat{H}_{\rm local}^2}-(\avg{\hat{H}_{\rm hopping}}+\avg{\hat{H}_{\rm local}})^2.
    \label{app:eq:DeltaE1}
\end{equation}
For our initial Fock state and Hamiltonian of 1d and 2d systems, $\hat{H}_{\rm local}$ does not change the state, merely counting the energies according to the excitations, leading to $\langle \hat H_{\rm local}^2\rangle = \langle \hat H_{\rm local}\rangle^2$. In turn, $\avg{\hat{H}_{\rm hopping}}=0$ as well is $\langle \hat H_{\rm hopping} \hat H_{\rm local}\rangle = 0$. Thus Eq.~(\ref{app:eq:DeltaE1}) results in
\begin{equation}
    \Delta E^2 = J^2_1N_1+J^2_2N_2,
\end{equation}
where $J_1$ ($J_2$) is the nearest (next-nearest) neighboring coupling and $N_1$ ($N_2$) is the corresponding number of hopping pairs. For our 1d case, $N_1=N-1$, where $N$ is the total number of qubits, and $\Delta E=|J_1|\sqrt{N_1-1}$ (see Fig.~\ref{smfig:compare_sizes}{\bf b}). For the 2d case, $\Delta E$ highly depends on the topology of the systems, and things are a little bit more complicated. For $2n\times2n$ systems, where $n\in \mathbb{Z}^+$, $N_1=2n(2n-1)$ and $N_2=2(2n-1)^2$. Thus one can have $\Delta E = \sqrt{2n(2n-1)J_1^2+2(2n-1)^2J_2^2}$. For $(2n+1)\times(2n+1)$ systems $N_1=4n(2n+1)$ and $N_2=0$. So $\Delta E=2|J_1|\sqrt{n(2n+1)}$. In summary, though its concrete form depends on the system's topology, $\Delta E$ is $W$-independent under our setup of the Fock initial state and checkerboard on-site potentials. For the initial states, which are superpositions of product states, the conclusion above will not be generically valid for $\avg{\hat{H}_{\rm hopping}}\neq0$ and $\Delta E$ can ultimately change with $W$.

Next, we will check if there is a connection between $\Delta E$ and the difference of the Fock energy with the crossover of MT and ML (ML$^{\star}$) bounds. Here, we assume the total number of qubits $N$ is twice the number of excitations $n$, $N=2n$, and $W$ is negative. The boundary between MT and ML regimes is identified by $E-E_{\rm min}=\Delta E$. Combining this with $\Delta E = 2W^*$, which means that the variance is exactly equal to the difference of consecutive Fock energies, one has that $E_{\rm min}=(n-2)W^*$. When the system is sufficiently large such that $n\gg2$, $E_{\rm min}\simeq nW^*$. The last paragraph shows that $\Delta E$ is proportional to the system size, and so is $W^*$. On the other hand, while the magnitude of the on-site potential is large, the hopping part of the Hamiltonian is comparably small to the local one, at which $E_{\rm min}\simeq nW$. This implies that when increasing the system size $W^* \simeq W$ and the crossover of MT and ML bounds occurs when the variance is equal to the difference of Fock energies, i.e., $\Delta E = 2W^*$ under our setup of the configurations of the Hamiltonian (see Fig.~\ref{smfig:compare_sizes}{\bf a}). Cases of odd $N$ and MT-ML$^{\star}$ are easily generalizable.       

\bibliography{sm}